\documentclass[a4paper, amsfonts, amssymb, amsmath, reprint, showkeys, nofootinbib, twoside]{revtex4-1}
\usepackage[english]{babel}
\usepackage[utf8]{inputenc}
\usepackage[colorinlistoftodos, color=green!40, prependcaption]{todonotes}
\usepackage{amsthm}
\usepackage{mathtools}
\usepackage{physics}
\usepackage{xcolor}
\usepackage{tikz}
\usepackage{graphicx}
\usepackage[left=23mm,right=13mm,top=35mm,columnsep=15pt]{geometry} 
\usepackage{cancel}
\usepackage{adjustbox}
\usepackage{placeins}
\usepackage[T1]{fontenc}
\usepackage{lipsum}
\usepackage{csquotes}

\makeatletter
\renewcommand{\fnum@figure}{\textbf{\figurename~\thefigure}}
\renewcommand{\fnum@table}{\textbf{\tablename~\thetable}}
\makeatother
\usepackage[percent]{overpic}
\usepackage{soul}
\usepackage{float}
\usepackage[pdftex, pdftitle={Article}, pdfauthor={Author}]{hyperref} 
\begin{document}

\title{\MakeUppercase{\fontsize{15}{15}\selectfont 
A field-biased quantum master \\ equation and its Markovian limit}}

\author{M. Gabriela Boada G.\textsuperscript{\textit{1}}}
    \email[Correspondence email address: ]{maria.boada@utsa.edu.} 
\author{Andrea Delgado\textsuperscript{\textit{3}}}
\author{Jose Morales E.\textsuperscript{\textit{1,2}}}
\affiliation{\textsuperscript{1}University of Texas, San Antonio, Dept. of Physics and A., San Antonio, TX 78249, USA.\\
\textsuperscript{2}University of Texas, San Antonio, Dept. of Mathematics, San Antonio, TX 78249, USA.}
\affiliation{\textsuperscript{3}Qblox Inc., Delft, 2628 CJ, The Netherlands.
}

\date{\today} 

\begin{abstract}
We present a non-equilibrium quantum master equation for a driven open quantum system in the presence of a continuously applied electromagnetic field. Starting from a driven Caldeira-Leggett (CL) model in which the external electromagnetic field couples simultaneously to the subsystem and reservoir degrees of freedom, the canonical fluctuation-dissipation theorem (FDT) relations that encode the coefficients of the master equation can no longer be expected to hold. The bath statistics acquire an explicit dependence on the two-time autocorrelation function of the applied field, leading to drive-biased noise correlations and the potential for non-Markovian dynamics. By eliminating the reservoir degrees of freedom at the operator level, we obtain a modified Hu-Paz-Zhang (HPZ) master equation, in which the diffusion coefficients and coherent forces inherit an explicit memory dependence on the external field. We demonstrate that the physically observable resonant frequency remains encoded in the homogeneous Green's function of the Generalized Langevin equation (GLE), while the drive-induced corrections manifest exclusively through modified diffusion and drift terms, should the drive be treated classically. The resultant field-modified HPZ master provides a unified microscopic framework for understanding field-biased open quantum systems with direct pertinence to a wide variety of experiments in quantum optics and microscopic quantum circuits.
\end{abstract}


\maketitle

\section{Introduction}

In the theory of open quantum systems \cite{BreuerPetruccione, GardinerZoller}, quantum master equations describe real-world interactions at the quantum mechanical scale \cite{FeynmanVernon1963}, wherein the environment inevitably exerts an influence over the sub-system, whose dynamics are the subject of interest, requiring both to be accounted for in the evolution of the relevant part \cite{Lindblad1976}. In the standard equilibrium sub-system/environment construction, the bath is prepared in a stationary finite-temperature thermal state completely characterized by time-translation–invariant two-time AutoCorrelation functions (AoC) \cite{FeynmanVernon1963}. Under these assumptions, the fluctuation–dissipation relations (FDT) from Quantum Field Theory (QFT) \cite{CallenWelton1951,Kubo,MartinSchwinger1959} establishes the universal correspondence between noise and dissipation, leading to a Markovian \cite{MartinSchwinger1959} master equation with coefficients that become stationary at the relevant system timescale, usually on the order of the system's thermal correlation time \cite{Lindblad1976,Gorini1976,Brasil2012}.

\begin{figure}
    \centering
    \includegraphics[width=.85\linewidth]{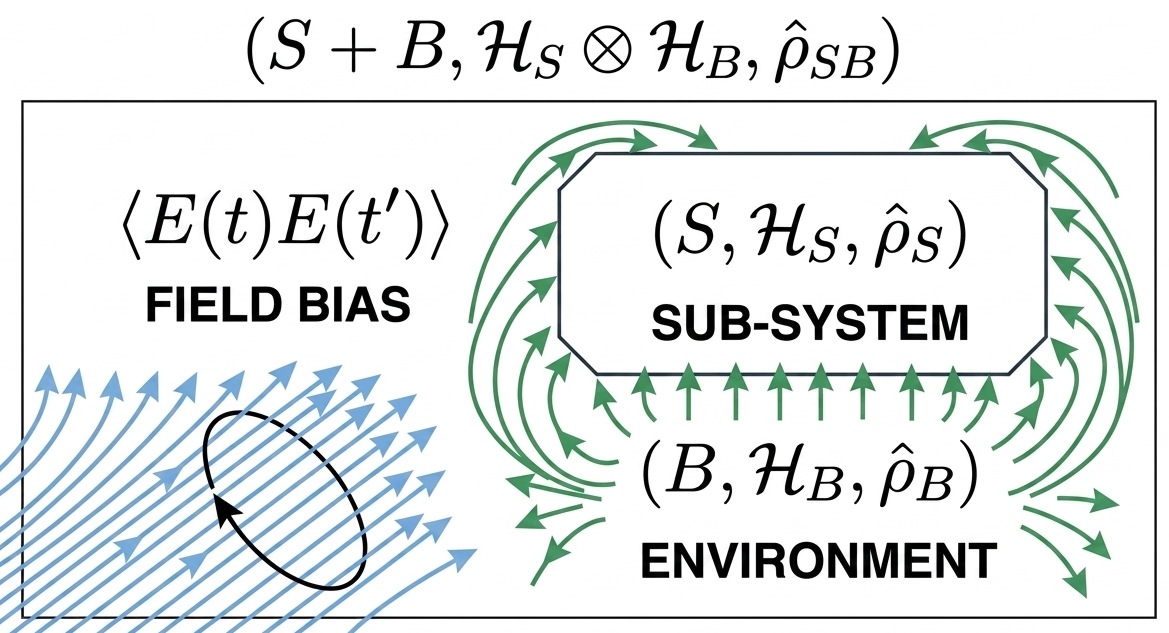}
    \caption{\textit{The field-biased open quantum system.} A schem- atic of the subsystem--environment construction used in the tagged particle--bath framework in \cite{CuiZaccone2018}. The total Hilbert space $S+B$ is decomposed into the principal subsystem $S$ with Hamiltonian $H_S$ and the environment $B$ with Hamiltonian $H_B$, coupling through the interaction Hamiltonian, $H_{SB}$, represented by the (green) arrows. The external, time-dependent electromagnetic field (blue), couples to both the sub-system and the resrvoir, while fringe fields are meant to represent erroneous drive-bath coupling.} 
    \label{fig: OQS_Schematic}
\end{figure}

For sub-systems driven continuously by an arbitrary electromagnetic field, $E(t)$, as depicted in \textbf{Figure \ref{fig: OQS_Schematic}.}, the equilibrium form of the FDT relations can no longer be assumed to hold arbitrarily \cite{PelargonioZaccone2023,CuiZaccone2018,GambaCuiZaccone2025}. Namely, 
the total bath force incurs drive-dependent correlations, and the canonical QFT-FDT relations \cite{Kubo},
\begin{align}
\label{eq: FDT}
\langle F(t) \rangle_{B} &=\; (\gamma e) E(t)  \nonumber \\  
\langle F(t)F(t') \rangle_{B}
&= mk_B T\nu(t-t') + \!(\gamma e)^2 E(t)E(t').
\end{align}
acquire an explicit non-stationary ($\langle F(t) \rangle \neq 0$) field bias dependence, directly proportional to the field's two-time AoC, $\langle E(t)E(t')\rangle_B$ \footnotemark. 
Furthermore, the quantum master equation presented here provides an accessible microscopic framework for describing both coherent interactions and field-biased noise on equal footing, with direct relevance to the control mechanisms in superconducting quantum circuits \cite{Blais2004,Wallraff2004,DevoretSchoelkopf2013}. 

One central question in driven-dissipative open quantum systems addressed here is therefore under what conditions the bias induced by the external field, such as a time-dependent RF drive, 
\begin{equation}
\label{eq: pump}
E(t) = 
\mathcal{E}_{P}(t)\cos(\omega_\mathrm{RF}\, t + \phi_\mathrm{RF}), \quad \mathcal{E}_{P}(t) \in \mathbb{C
}^\mathrm{1d}
\end{equation}
can be safely overlooked.~%
It can thus be shown that even when the reduced sub-system dynamics would admit a Markovian limit, the drive-induced AoC
\begin{align}
E(t)E(t')
&=
\frac{\mathcal{E}_{P}(t)\mathcal{E}_{P}(t')}{2}\cos(\omega_\mathrm{RF}(t-t'))
\\&+
\frac{\mathcal{E}_{P}(t)\mathcal{E}_{P}(t')}{2}\cos(\,\omega_\mathrm{RF}(t+t')+2\phi_\mathrm{RF}).
\end{align}
can generate cross-memory effects to the kernels which encode the coefficients of the master equation at the operator level. It is therefore desirable to formulate the criteria, expressed directly in terms of $\langle E(t)E(t')\rangle_B$, that determine when the equilibrium approximation remains valid and when drive-induced nonstationarity must be retained to account for the effects of drive-induced thermalization accurately.\setcounter{footnote}{1}\footnotetext{Denoting \(\langle \cdot \rangle_E \equiv \mathrm{Tr}_E[\cdot\, \rho_E(0)]\)}

In this work, we'll address this problem by means of a field-bias modification to the Hu–Paz–Zhang (HPZ) quantum master equation \cite{HuPazZhang1992}. The HPZ equation describes the exact reduced dynamics of a sub-system linearly coupled to a harmonic reservoir ensemble with frequency-biased noise, and reduces to the familiar local (Markovian) Caldeira-Leggett (CL) quantum master equation for Quantum Brownian Motion (QBM) in the high-temperature, weak-coupling limit of Ohmic dissipation \cite{CaldeiraLeggett1983,BreuerPetruccione}. Here, we extend this treatment to the case in which an external classical or quantum-mechanical field simultaneously drives the system and introduces a bias in the reservoir, yielding diffusion and drift coefficients that depend explicitly on the bias. 

While previous works have derived the generalized Langevin equation (GLE) for driven systems and their corresponding modifications to the FDT relations \cite{CuiZaccone2018,PelargonioZaccone2023,GambaCuiZaccone2025}, the systematic formulation of a non-Markovian master equation with explicitly field-dependent coefficients has not been presented before. Thus, our modification to the HPZ provides a generalized alternative directly accessible to the language of driven-dissipative quantum circuits \cite{Blais2004,Wallraff2004,DevoretSchoelkopf2013}, accounting for the renormalization of the physically observable frequency and the modification of the coefficients of the quantum master equation in the presence of a field bias.

In particular, the structure of the coefficients appearing in the master equation appears to suggest that ordinary Markovian models may be ill-suited for vastly multiplexed (or many-body) systems commonly encountered in large-scale superconducting processors \cite{Blais2021,Krantz2019,Gambetta2017,Arute2019,Walter2017} and/or Raman-assisted mixing schemes \cite{Jeff, Shruti, Mundhada2019} which require the application of many off-resonant pumps to be present throughout the duration of the protocol, since the modification to the equilibrium prefactor ultimately results in the rescaling $\propto \langle E(t) E(t')\rangle$, which does not ordinarily decay without intervention at the hardware level. Thus, our field-biased quantum master equation provides a generalized alternative directly accessible to the language of quantum circuits, accounting for the renormalization of the physically observable frequency and the modification of the coefficients of the quantum master equation in the presence of biasing fields, particularly suited to study scaled processor noise.

\section{Background}

Following \cite{CaldeiraLeggett1983,BreuerPetruccione}, we consider a sub-system associated
with bare frequency $\omega_S$, mass $m_S$ 
and position $x$ under linear coupling to the canonical CL reservoir ensemble, as in \cite{CuiZaccone2018}. The combined Hamiltonian is 
\begin{equation}
H_{SB}=H_S + H_B + H_I
\end{equation} 
where
\begin{align}
H_S &=\frac{p^2}{2m_S}
+
\frac{m_S\omega_S^2}{2}  x^2 - xE(t),
\\
H_B &= 
\sum_j
\Big[\;+
\dfrac{1}{2}\frac{p_j^2}{m_j}
+
\frac{m_j}{2} \omega_j^2 x_j^2  \;\Big],
\\
H_I &=\sum_j
\Big[
-
d_j x_j E(t)
+
c_j x_j x \,\Big],
\label{eq:driven_CL_H}
\end{align}
where $\{x_j\}$ and $\{p_j\}$ are the reservoir oscillator position and momentum coordinates, $\{\omega_j\}$ are their frequencies, and $\{c_j\}$ are the system--bath coupling constants while $\{d_j\}$ are the drive--bath coupling rates. 

 The mechanical quantities appearing in the CL model can be expressed in terms of inductive/capacitive circuit elements by promoting the position coordinate to the node flux $x \longrightarrow \hat{\varphi}$ and the momentum variable with the conjugate charge $p\longrightarrow  \hat{\varphi}_q$ \cite{bishop2010cqed}. The mass $m_S$ is replaced by the lumped circuit \cite{BBQ} capacitance $C_S$, while the frequency $\omega_{S}$ is naturally $\omega_{S} = 1/\sqrt{L_{S}C_{S}}$. In this map, $H_S \longrightarrow q^2/(2C_{S}) + \varphi^2/(2L_{S}) - \varphi E(t)$. 
 
 The coupling constants \( c_j \) arise from capacitive interactions between the principal node and each bath \enquote{resonator} mode \cite{blais2021circuitqed}, modeled as a microwave (RF) transmission line \cite{pozar2012microwave}. Then, one identifies \( c_j \sim C_{c,j} \, V_{\text{zpf,}j} \), where \( C_{c,j} \) is the mutual coupling capacitance and \( V_{\text{zpf,}j} \) is the zero-point voltage fluctuation for the \( j \)-th mode, which leads to the system--bath coupling rate as (sometimes denoted $g_{SB}$) 
\begin{equation}\gamma_{S}\sim\sum_j\frac{C_{cj}}{\sqrt{C_s C_j}}\sqrt{\omega_s\omega_j}\end{equation}
and bath--field coupling rate 
\begin{equation}\gamma_{E}\sim C_{Ej}V_{\mathrm{zpf}^j}\mathcal{E}_{P}(t).\end{equation}

The quantity $\mathcal{E}_{P}(t):=(\mathcal{E}_{\mathrm{RF}}\hbar)^2$ fixes the characteristic energy scale of the time-dependent drive, such that the field amplitude $\mathcal{E}_{P}(t)$ is expressed in units of power (e.g., $\mathrm{V}$ or $\mathrm{V/m}$, depending on the normalization). The drive center frequency, $\omega_{\mathrm{RF}}$, has units of ($\mathrm{rad,\;s^{-1}}$), and $\phi_{\mathrm{RF}}$ is a dimensionless phase offset.
The external drive enters as a classical voltage bias applied through an additional capacitance at the ground pads $c_\mathrm{d}\equiv C_\mathrm{in}/C_\mathrm{out}$. 


\begin{table}[H]
\label{tab:experimental_parameters}
\centering
\begin{ruledtabular}
\begin{tabular}{lcc}
Parameter & Range & Chosen \\
\hline
\( \omega_{S}/2\pi \) 
& \( 4.0\text{--}8.0 \,\mathrm{GHz} \) 
& \( 5.0 \,\mathrm{GHz} \) \\
\( \gamma_{B}/2\pi \) 
& \( 1.0\text{--}50 \,\mathrm{MHz} \) 
& \( 10 \,\mathrm{MHz} \) \\
\( \gamma_{E}^\mathrm{(s)}/2\pi \) 
& \( 0.1\text{--}20 \,\mathrm{MHz} \) 
& \( 2.0 \,\mathrm{MHz} \) \\
$\Omega_{B}/2\pi$ &  $1.0$--$20.0~\mathrm{GHz}$ & $8.0~\mathrm{GHz}$ \\
\( \gamma_{E}/2\pi \) 
& \( 0.1\text{--}10 \,\mathrm{MHz} \) 
& \( 1.0 \,\mathrm{MHz} \) \\
\end{tabular}
\end{ruledtabular}
\caption{\textit{Physical circuit parameters.} \footnotesize
Illustrative parameter set for a quantum-mechanical treatment of the drive, with corresponding circuit-level quantities. 
The chosen ranges are consistent with typical linear components in circuit QED architectures: superconducting resonator frequencies in the $4$--$8~\mathrm{GHz}$ range are standard for transmon-based devices \cite{blais2021circuitqed,strongcoupling_blais2004}, 
while capacitive coupling strengths in the $1$--$50~\mathrm{MHz}$ range arise from experimentally realizable coupling capacitances and participation ratios \cite{koch2007transmon,paik2011observation}. 
Drive and environment coupling rates in the $0.1$--$20~\mathrm{MHz}$ range reflect typical external line coupling and control amplitudes used in microwave-driven experiments \cite{reagor2016quantum,gertler2021protecting}. The cutoff frequency $\lambda$ in the GHz range reflects typical environmental mode bandwidths and engineered filter scales in circuit QED \cite{strongcoupling_blais2004}.} 
\end{table}

\subsection{Generalized Langevin Equation}

To determine the time evolution of the relevant subsystem under the influence of the structured reservoir, we can eliminate the reservoir degrees of freedom at the operator level \cite{halliwell1996alternative}. The Heisenberg Equations of Motion (EoM) \cite{heisenberg1925, born1925} for the bath oscillators follow directly from the total Hamiltonian as defined in \textbf{Eq.~(\ref{eq:driven_CL_H})}. Each bath oscillator satisfies a driven harmonic oscillator equation \cite{halliwell1996alternative},
$
m_j \ddot x_j(t) + m_j \omega_j^2 x_j(t) = c_j x(t) + E(t)
$, %
which can be written exactly in terms of the  $j$-th solution to the homogeneous problem, denoted as 
\begin{align}
\label{eq:qj_solution}
x_j(t)
&=
x_j^{(h)}(t)
\\&+
\int_{0}^{t} ds\, G_j(t-s)
\Big[\,
c_j x(s) + E(s)
\Big],
\label{eq:qj_solution_kernel}
\end{align}
where $G_j(t)$ is the inhomogeneous Green's function for the $j$-th bath oscillator. Substituting \textbf{Eq.~(\ref{eq:qj_solution})} into the Heisenberg EoM, 
\begin{align}\nonumber
m_S\ddot x(t) &+ m_S\omega^2 x(t)
=
\sum_j c_j x_j^{(h)}(t)
\\&\quad\quad\quad+
\sum_j c_j^2 \int_{0}^{t} ds\, G_j(t-s)x(s)
\\&\quad\quad\quad
+
\sum_j c_j \int_{0}^{t} ds\, G_j(t-s)E(s).
\label{eq: LangevinBare}
\end{align}
The first term in the RHS of \textbf{Eq. (\ref{eq: LangevinBare})}  represents the fluctuating force, the second term thus describes the dissipative back-action. Collecting terms,
\begin{equation}
m_S(\ddot x(t)
+ \omega^2 x(t)) +
\int_{0}^{t} ds\, \gamma(t-s)\dot x(s)
=
F(t),
\label{eq:driven_GLE}
\end{equation} 
results in the driven GLE \cite{CuiZaccone2018}. The total force correlator $F(t)$ separates naturally into two contributions, 
\begin{equation}
\label{eq: totalforce}
F(t) = F^{\mathrm{eq}}_{B}(t) + F_E(t)\end{equation}
defined by
\begin{align}F_{B}(t)&:= \sum_j c_j x_j^{(h)}(t), \\ \label{eq: fieldcorrelator}
F_E(t)&:=\sum_j c_j \int_{0}^{t} ds\, G_j(t-s)E(s).\end{align}
$F(t) $ originates from the equilibrium bath force, whereas $F_E(t)$ represents the deterministic response induced by the external field. The external field drives each mode, and its response at the time \( t \)
\begin{equation}
\label{eq: memorykernel_general}
\gamma(t) = \sum_j c_j^2 \int_0^t ds \, G_j(t - s)\, x_j(s),
\end{equation}
is the accumulated effect of the past drive \( G_j(s) \) as in \textbf{Eq.~(\ref{eq:qj_solution})-(\ref{eq:qj_solution_kernel})}, leading to the homogeneous solution $x_j(t)$ appearing naturally as the causal propagator that maps past forcing into present response. Thus, the so-called memory kernel $\gamma(t)$ can be expressed directly
by the substitution of the homogeneous solution, $x(t) \rightarrow x^{(h)}(t)$ and $G_j(t)$ into \textbf{Eq. (\ref{eq: memorykernel_general})} along the characteristics of the initial conditions of the GLE. 

The remaining statistical properties of the subsystem are uniquely determined by the total force correlator appearing in the GLE and the equilibrium bath specification. In the following sections, we show how this drive-modified memory kernel $\gamma(t)$ and total noise correlator $F(t)$ determine the dynamics of the reduced subsystem's evolution, and how the bath specification is made and passed to the continuum limit. 

\subsection{Fluctuation-Dissipation theorem}

While single-time expectation values characterize instantaneous properties of a stochastic (or quantum mechanical) process, the dynamical influence of the $j$-th environmental mode over the sub-system is fundamentally encoded in its two-time AoC (see: \cite{BreuerPetruccione} pp. 134-135),
\begin{align}
\langle F_{j}^{\dagger}(t),F_{k}(t-s) \rangle_B &\equiv  
\mathrm{Tr}_B\big[F_{j}^{\dagger}(t) F_{k}(t-s) \rho_B\big] 
\end{align}
wherein we utilize the shorthand $t'\equiv t-s$ on $(t \geq s \geq 0)$ throughout. These mathematical objects readily quantify how fluctuations at one time influence the sub-system's evolution at a later time, and then provide the descriptions for emergent non-Markovian behavior. 

In particular, stationarity \cite{breuer2002theory} is recovered when the quantity retains or gains time-translation invariance on the order of the relevant sub-system timescale \cite{Kubo}. For instance, let $\rho_B$ be a stationary thermal state \cite{gibbs1902} of the reservoir at $t=0$. In the interaction picture \cite{heisenberg1925}, the bath Hamiltonian commutes, meaning  $[H_B, \rho_B]=0$, leading to 
\begin{align}
\label{eq:AoC1}
    \langle F_j^{\dagger}(t) F_{k}(s) \rangle_B &= 0,\quad \forall\, j,\, k,\\ \label{eq:AoC2}
    \langle F_j(t) F_{k}(s) \rangle_B &= 0 = \langle F_j^{\dagger}(t) F_k^{\dagger}(s) \rangle_B,
\end{align}
resulting in 
\begin{align}
\langle F_{j}^{\dagger}(t),F_{k}(t-s) \rangle_B &\equiv  
\langle F_{j}^{\dagger}(s),F_{k}(0) \rangle_B
\end{align}
which leads to frequency-dependent coupling rates that do not change over time. 

In standard treatments, the total force correlator appearing in the driven GLE is decomposed into its so-called symmetric and anti-symmetric parts, denoted 
\begin{align}
\label{eq: symmetric}
\frac12\langle \big[\,F(t),F(t')\,\big]\rangle_{B}
&=
\nu_F(t,t'),
\\
\frac{1}{2i}\langle \{F(t),F(t')\} \rangle_{B}
&=
\mu_F(t,t'),
\end{align}
respectively. The antisymmetric correlator, also known as the noise kernel, 
describes the fluctuations of the force correlator acting on the subsystem and therefore determines the diffusive response.  Similarly, the symmetric part
characterizes the dissipative response of the environmental modes and is directly related to the retarded susceptibility of the bath force operator. To understand why, it is illustrative to take a note of the correspondence between the time-domain kernels introduced above in the language of response functions.

\subsubsection{Applied theory of response functions}

The decomposition of the reservoir AoC into dissipative and fluctuation contributions is most transparently understood at the level of their symmetry properties in time. In the frequency domain, the symmetric correlator produces a real-valued two-sided noise power spectrum,
\begin{equation}
S_{F}(\omega,\omega')
=
\int_{-\infty}^{+\infty} dt
\int_{-\infty}^{+\infty} dt'\,
e^{i\omega t - i\omega' t'}
\nu_B(t,t').
\end{equation}
The two-time causal response function is built from the antisymmetric correlator:
\begin{equation}
\chi_F(\omega,\omega')
=
\frac{i}{\hbar}
\int_{-\infty}^{+\infty} dt
\int_{-\infty}^{+\infty} dt'\,
e^{i\omega t - i\omega' t'}
\tilde{\chi}_F(t,t').
\end{equation}
where we have introduced the explicit two-time retarded (causal) response kernel
\begin{equation}
\tilde{\chi}_F(t,t')
=
\frac{i}{\hbar}\,
H(t - t')\,
\mu_F(t,t') ,
\end{equation}
whose imaginary part governs irreversible energy flow into the reservoir \cite{Kubo}. The Heaviside function $H(t - t')$ enforces causality in the quantum mechanical cognate of the response function. In particular, it implies that the retarded kernel satisfies $\Tilde{\chi}_F(t,t') = 0$ for $t < t'$. Physically, this condition ensures that the system's response at time $t$ depends only on earlier times $t'$, so that future perturbations cannot influence the past.  With these definitions in place, the standard stationary quantities introduced are recovered as the special case in which the bath correlations become time-translation invariant, and the two-time spectra collapse onto their single-time diagonal (stationary) components. 

The FDT relates the Fourier transforms of the abovementioned
quantities through
\begin{equation}
\label{eq: eqNoisePower}
S_{FF}(\omega)
=
\frac{\hbar}{2}\coth\!\left(\frac{\hbar\omega}{2k_B T}\right)
\big[\chi_F(\omega)-\chi_F^*(\omega)\big].
\end{equation}  
For a reservoir in thermal equilibrium at $T$, this relation drastically simplifies because
$$\chi_F(\omega)-\chi_F^*(\omega)=2i\,\mathrm{Im}\,\{\chi_F(\omega)\}$$ yields the equilibrium noise power spectrum. %
In particular, $\operatorname{Im}\chi_F(\omega)$ corresponds to the absorptive component of the response and therefore encodes dissipation, while the real part describes the dispersive (reactive) response. Thus, fluctuations and dissipation are not associated with the imaginary and real parts of the same object, but rather arise from distinct correlation functions: the symmetrized correlator determines the noise spectrum, whereas the commutator determines the causal response. In thermal equilibrium, these quantities are related by the FDT, but this connection is lost in the presence of external driving, where both kernels must be treated independently as two-time functions.

Equivalently, introducing the center and relative coordinates, $T = (t+t')/2$ and $\tau = t-t'$, one may define the alternate representation in the shifted coordinate space (in the likeness of the Wigner function)
\begin{align}
\label{eq: shiftedspace}
S_{FF}(\omega;T)
&=
\int_{-\infty}^{+\infty} d\tau\,
e^{i\omega \tau}\,
\nu_F\!\left(T+\frac{\tau}{2},\,T-\frac{\tau}{2}\right),
\end{align}
which makes explicit how the instantaneous noise power spectrum evolves in time under the influence of the drive. The double-frequency representation $S_{FF}(\omega,\omega')$ is then obtained as the Fourier transform of $S_{FF}(\omega;T)$ with respect to the center time $T$, providing a complete description of nonstationary fluctuations. 

The appearance of two independent frequencies has a clear physical interpretation. For a stationary reservoir, correlations depend only on the time difference $(t - t')$, and their corresponding Fourier transform is diagonal, $S_{FF}(\omega,\omega') \propto \delta(\omega-\omega')$, expressing energy conservation and the absence of frequency mixing. In contrast, when the reservoir is driven, correlations depend separately on the central time and relative time, and the spectrum acquires off-diagonal components. These encode frequency conversion processes mediated by the drive, whereby fluctuations at one frequency $\omega'$ can contribute to the system dynamics at a different frequency $\omega$.

\subsubsection{General noise and dissipation kernels}

Without explicitly assuming stationarity of the total bath correlator, the noise kernel \(\nu(t,t')\) and the response kernel \(\chi_B(t,t')\) are no longer connected by the classical equilibrium FDT, but instead inherit a common dependence on the field through the driven bath correlators. It should be clear through the direct insertion of the total force correlator into these definitions that the kernels acquire additional non-stationary contributions under the field bias, denoted as 
\begin{equation}
\label{eq: noisekernel}
\nu(t,t')
=
\nu^{\mathrm{eq}}_{B}(t-t')
+
\nu_E(t,t')
+
\nu^{\mathrm{neq}}_{EB}(t,t'),
\end{equation}
and 
\begin{equation}
\mu(t,t')
=
\mu^{\mathrm{eq}}_{B}(t-t')
+
\mu_E(t,t')+
\mu^{\mathrm{neq}}_{EB}(t,t').
\end{equation}
Physically, this is a reflection of the fact that even under the classical treatment, the drive injects energy into the system and produces additional fluctuations in the total force, though it does not alter the intrinsic quantum fluctuations of the bath or its dissipative response under the classical treatment. Expanding the commutator in the anti-symmetric correlator, one may identify
\begin{align}
\nu_B^{\mathrm{eq}}(t-t')
&=
\frac{1}{2i}
\left\langle
\left[
F_B(t),F_B(t')
\right]
\right\rangle ,
\\
\nu_E(t,t')
&=
\frac{1}{2i}
\left\langle
\left[
F_E(t),F_E(t')
\right]
\right\rangle ,
\\
\nu_{EB}^{\mathrm{neq}}(t,t')
&=
\frac{1}{2i}
\langle
\left[
F_B(t),F_E(t')
\right]
\\&\quad\quad\quad+
\left[
F_E(t),F_B(t')
\right]
\rangle .
\end{align}

Similarly, expanding the symmetric correlator gives
\begin{align}
\mu_B^{\mathrm{eq}}(t-t')
&=
\frac{1}{2}
\left\langle
\left\{
F_B(t),F_B(t')
\right\}
\right\rangle ,
\\
\mu_E(t,t')
&=
\frac{1}{2}
\left\langle
\left\{
F_E(t),F_E(t')
\right\}
\right\rangle ,
\\
\mu_{EB}^{\mathrm{neq}}(t,t')
&=
\frac{1}{2}
\langle
\left\{
F_B(t),F_E(t')
\right\}
\\&\quad\quad+
\left\{
F_E(t),F_B(t')
\right\}
\rangle .
\end{align}

For the quantum treatment of the drive, the field-induced force is no longer a
$c$-number contribution. Instead, the total force operator is written as
$
F(t)=F_B(t)+F_E(t),
$
where $F_B(t)$ is the intrinsic bath force and $F_E(t)$ is the operator-valued force
generated by the quantum electromagnetic field. 

The key distinction from the classical-drive treatment is that $F_E(t)$ is now an operator rather than a deterministic function. Consequently, the field-field term $\nu_E(t,t')$ is a genuine quantum noise contribution rather than simply $F_E(t)F_E(t')$, and the field commutator term $\mu_E(t,t')$ can contribute to the
dissipative kernel. In addition, the cross terms $\nu_{EB}^{\mathrm{neq}}(t,t')$ and $\mu_{EB}^{\mathrm{neq}}(t,t')$ do not vanish when the bath and field are correlated by their mutual coupling rate.

\subsection{The continuum limit description}

\begin{figure}
    \centering
    \includegraphics[width=\linewidth]{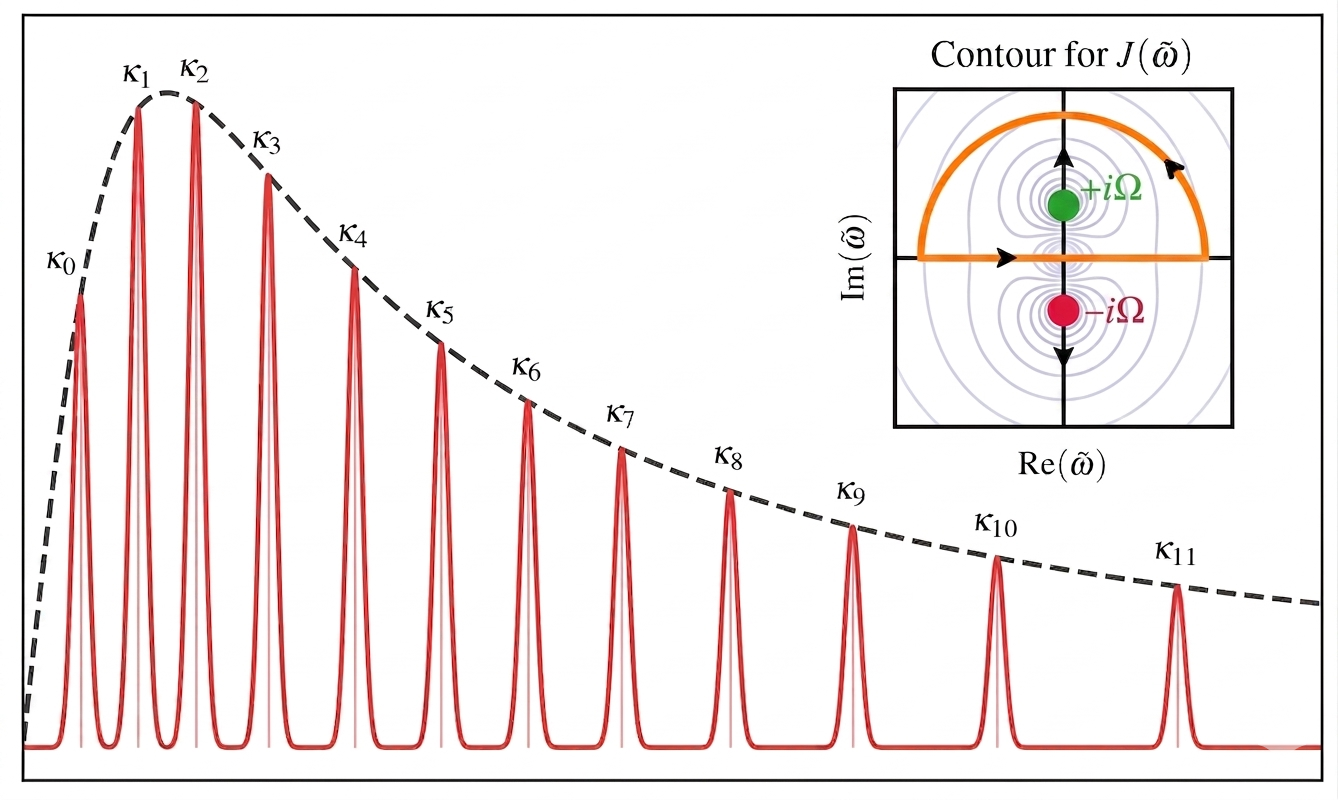}
    \caption{\textit{Continuum-limit of Ohmic dissipation.} A depiction of the continuum limit of Ohmic dissipation for the sparsely spaced set of modes to illustrate the nature of the approximation. Frequency dependent couplings from the discrete set of reservoir oscillators are replaced by the corresponding Dirac-Delta functions. In the subceptibility domain, shown in the inset, with the first order poles $\pm \Omega$ for the Ohmic response function. The positive pole is by a semi-circle, isolating the converging $t\geq 0$ region utilized to evaluate the expressions for the kernels using residue theorem.}
    \label{fig:continuum}
\end{figure}

The preceding expressions were written for a discrete set of bath oscillator modes. In physical environments, the bath typically consists of a very large number of closely spaced modes (see: \textbf{Figure \ref{fig:continuum}}), and it becomes convenient to replace the discrete sums over reservoir indices by a continuum-limit description \cite{BreuerPetruccione} written in terms of the spectral density
\begin{equation}
J(\omega)
=
\frac{\pi}{2}
\frac{\kappa_j^2}{m_j\omega_j}
\,
\delta(\omega-\omega_j).
\end{equation}

In this representation, the noise kernel (strictly under the classical treatment of the drive) introduced in becomes
\begin{align}
\gamma(t)
&= \sum_j \frac{c_j^2}{m_j\omega_j^2}\cos(\omega_j t) \\&\simeq 
\frac{2}{\pi}
\int_0^\infty d\omega
\,
\frac{J(\omega)}{\omega}
\cos(\omega t),
\end{align}

The continuum description applies when the field possesses a finite spectral bandwidth and can be modeled as a densely packed continuum \cite{BreuerPetruccione, CaldeiraLeggett}. 
Invoking the definition of the linear response of the reservoir oscillators, one can see that 
$
\operatorname{Im}\{\chi_F(\omega)\}=J_{B}(\omega),
$
so that the equilibrium noise power spectrum in \textbf{Eq. (\ref{eq: eqNoisePower})} may be written directly in terms of
the equilibrium bath spectral density $J_{B}(\omega)$. The equilibrium kernels assume the well-known time-invariant form,
\begin{align}
\label{eq: equilibrium_noise}
\nu^{\mathrm{eq}}_{B}(t-t')
&=
\frac{1}{\pi}
\int_0^\infty d\omega\,
J_{B}(\omega)
\sin[\,\omega(t-t')\,]. \\ 
\mu^{\mathrm{eq}}_{B}(t-t')
&=
\frac{1}{\pi}
\int_0^\infty d\omega\,
J_{B}(\omega)
\nu_T 
\cos[\,\omega(t-t')\,].
\end{align}
where 
\begin{equation}
    \nu_T = \coth\!\left(
\frac{\hbar\omega}{2k_B T}
\right).
\end{equation}
The appearance of the thermal factor $\nu_T$ admits a natural Matsubara decomposition \cite{matsubara1955}, in which the noise kernel is expressed as a sum over the discrete imaginary frequencies corresponding to the poles of the Bose–Einstein distribution. This representation is particularly useful because it converts the continuum integral into a sum of exponentially decaying contributions in time, making the memory structure explicit and enabling efficient numerical treatments of the kernel.

Importantly, the continuum limit is where the bath specification is made, and while not particularly necessary for a classical treatment of the field-bias, the spectral description in \textbf{Eq. (\ref{eq:  shiftedspace})} can still be made to pass the field to the continuum limit. In the sub-sections to follow, we make the bath specification for regularized Ohmic dissipation under the monochromatic field bias.

\subsubsection{The bath specification: \\ 
On Ohmic-Lorentz Drude dissipation}

A commonly adopted phenomenological model for an equilibrium bath is the Ohmic Lorentz-Drude spectral density, characterized by a linear dependence in the low-frequency limit, $J_{B}(\omega) \propto \omega,$ where $(\omega \to 0)$. Introducing the frequency-independent damping coefficient $\gamma_B$, one specifies the spectral density $J_{B}(\omega)\rightarrow J_\mathrm{Ohm}(\omega)$, 
\begin{equation}
    J_B(\omega) = \frac{2 m \gamma_B}{\pi}\,\omega,
\end{equation}
which yields 
a friction kernel corresponding to memoryless (Markovian) coupling at a damping rate $\gamma_B$. This form captures the regime in which the bath response is proportional to velocity, reproducing classical viscous dissipation within a quantum description. 

However, such a strictly linear spectrum is unphysical at high frequencies \cite{CaldeiraLeggett1983}, since it can lead to divergent renormalizations of the system. To regularize the model, one introduces a high-frequency cutoff frequency $\Omega_B$, most commonly through a Lorentz--Drude form,
\begin{equation}
    \label{eq: Lorentz}
    J_{B}(\omega) = \frac{2 m \gamma_{B}}{\pi}\,\omega\,\frac{\Omega_{B}^2}{\omega_{B}^2+\Omega_{B}^2}.
\end{equation}
This modification preserves the Ohmic behavior at low frequencies while suppressing the contribution of high-frequency modes, thereby rendering both the dissipation kernel and the frequency renormalization finite. In the time domain, the cutoff introduces a finite bath correlation time $\tau_B \sim \Omega^{-1}_{B}$.

\subsubsection{The field specification: \\ On coherent and incoherent drive forces}

\begin{figure}
    \centering
    \includegraphics[width=\linewidth]{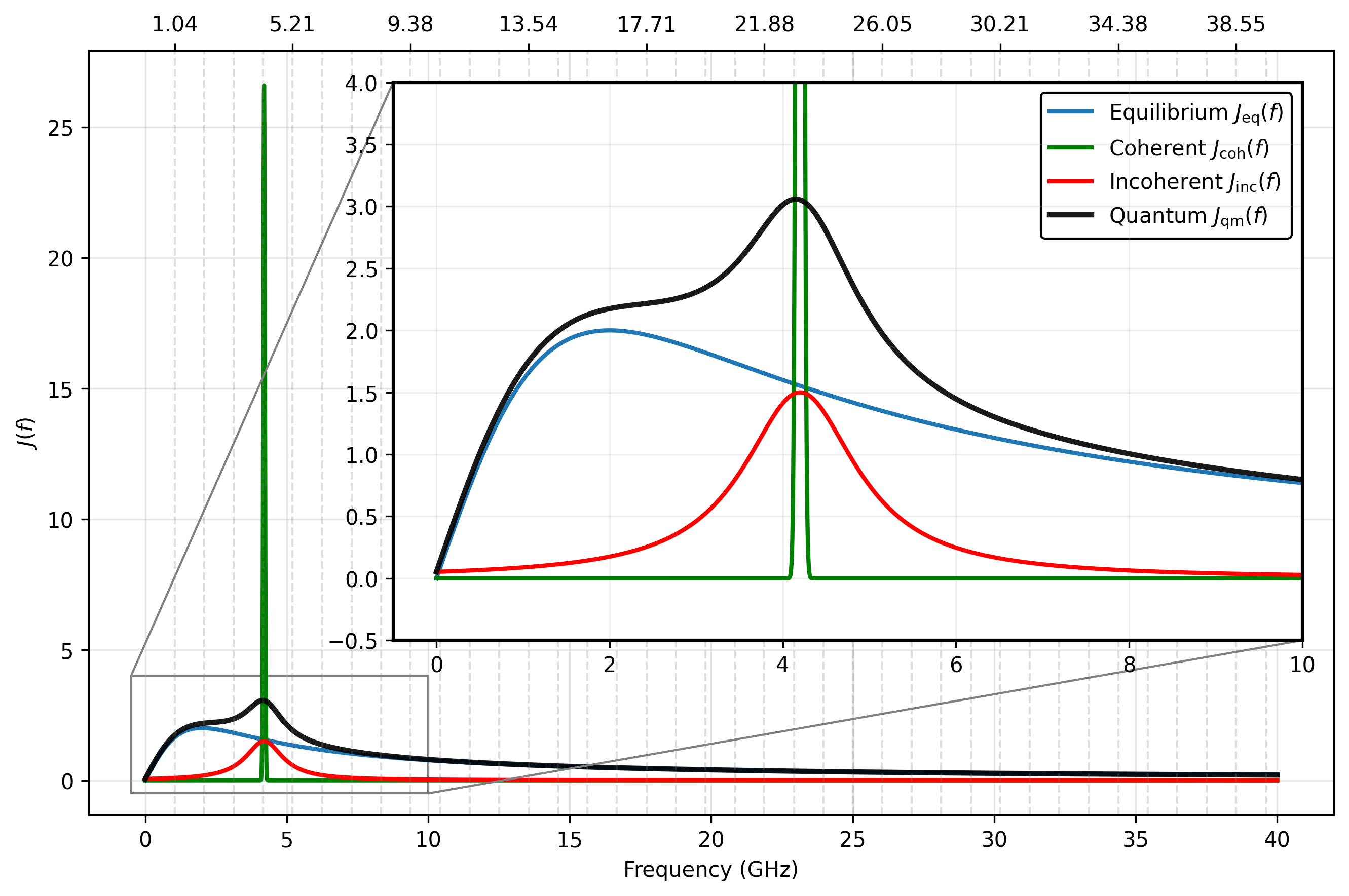}
    \caption{\textit{Coherent \& incoherent monochromatic field spectrum.} A decomposition of the effective bath/field coupling spectrum $J(\omega)$ (black) into its constituent contributions. The equilibrium bath spectral density $J_{\mathrm{eq}}(\omega)$ (blue) provides a smooth broadband background, while the coherent drive induces a narrow, delta-like peak $J_{\mathrm{coh}}(\omega)$ (green) centered at the drive frequency. In addition, the field--bath interaction generates an incoherent continuum contribution $J_{\mathrm{inc}}(\omega)$ (red), arising from cross-correlations between the drive and bath degrees of freedom, with markings and labels on the top axis at select Matsubara frequencies. The resulting total spectral density exhibits both a sharp resonant feature and a modified broadband structure, as highlighted in the inset.}
    \label{fig:coherentdrive}
\end{figure}

An instantaneous quantum mechanical drive superimposes a sharply localized spectral feature on the intrinsic reservoir background (see: \textbf{Figure \ref{fig:coherentdrive}}), producing an incoherent contribution through cross-bath correlations in addition to a direct, coherent displacement. The incoherent contribution modifies the fluctuations in the subsystem, consistent with the drive injecting energy into the system, by introducing a periodic forcing term at the operator level of the GLE.

Assuming the regularized Ohmic Lorentz-Drude spectral density for the equilibrium background frequency-dependent couplings as in the previous sub-section, the spectral density associated with the coherent pump filtered through the field-bath coupling is
\begin{align}
J_{E}(\omega)
&= \frac{\pi \gamma_{E}^{2} A_\mathrm{RF}^{2}}{2}\,\delta(\omega+\omega_\mathrm{RF})
\\&+ \frac{\pi \gamma_{E}^{2} A_\mathrm{RF}^{2}}{2}\,\delta(\omega-\omega_\mathrm{RF})
\end{align}
while the cross-correlations result in 
\begin{align}
J_{E}^{B}(\omega)
=
\frac{A_{\rm RF}^{2}}{4\pi}
\Big[&
\frac{\gamma_E}
{(\omega+\omega_{\rm RF})^2+(\gamma_E/2)^2}+
\\&\quad
\frac{\gamma_E}
{(\omega-\omega_{\rm RF})^2+(\gamma_E/2)^2}
\Big].
\end{align}
The coherent contribution $\propto \delta(\omega\pm\omega_\mathrm{RF})$ does not represent noise, but rather the direct coherent (deterministic) displacement encoded at the level of two-time correlation functions, demonstrating that the coherent spectral component acts through the system susceptibility rather than as a contribution to the fluctuations. 

As we will discuss throughout the remainder of the text, this establishes that the coherent monochromatic field may be equivalently incorporated either as an explicit forcing term $F_E(t)$ in the GLE, or a nonstationary contribution $F_E(t)F_E(t')$ from the uncentered bath autocorrelator, but not both. 

\section{Quantum Master Equation}

By invoking the CL reservoir \cite{caldeira1983path, CaldeiraLeggett}, we have implicitly assumed that the bath remains Gaussian and factorizes under field-bias, hence the reduced subsystem dynamics can be obtained exactly utilizing the same methods employed through an alternative derivation of the HPZ \cite{halliwell1996alternative}. The method lies on closure to the classical limit in Wigner space \cite{Pedestrians} to recast the operator-valued master equation as a c-number evolution equation in phase space  whose characteristics coincide with those of the GLE, thus providing the direct bridge between quantum dissipation as described by the master equation and classical stochastic motion as described by the GLE. This method is particularly attractive in the context of superconducting circuits, where experimentally accessible observables are naturally interpreted in terms of classical stochastic EoM, allowing the GLE description to serve as an intuitive (and quantitatively faithful) bridge between microscopic quantum models and measured device behavior (see: \cite{LongQuote} on pp. 134 in \cite{Blais2021}).

Following the same correspondences, the HPZ quantum master equation (and its phase-space cognates) can be \textit{modified} by \textit{factorizing} the drive forcing term
\begin{align}
\frac{\partial W}{\partial t}&=
\{H_{\mathrm{ren}}(t),W\}_{\mathrm{MB}}
+ \mathcal{F}_E^{(W)}(t)W \\&+2\hbar \,\Gamma(t)\frac{\partial}{\partial p}\bigl[pW\bigr] 
\\&+\hbar^2 D_{pp}(t)\frac{\partial^2 W}{\partial p^2}-\hbar^2 D_{xp}(t)\frac{\partial^2 W}{\partial x\,\partial p}
\end{align}

Here \(\{H_{\mathrm{ren}},W\}_{\mathrm{MB}}\) is the Moyal bracket \cite{moyal1949},

\begin{align}
\label{eq: moyal}
\{H_{\mathrm{ren}}(t),W\}_{\mathrm{MB}}
&=
-\frac{p}{m}\frac{\partial W}{\partial x}
+
m\Omega_{\mathrm{ren}}^2 x\frac{\partial W}{\partial p}
\end{align}
which, for a quadratic Hamiltonian,
reduces to the Poisson bracket \cite{Poisson} appearing in the classical Liouville \cite{liouville1838}. 

An important consequence of the specified drive-bath coupling is the renormalization of the physically
observable frequency arising from the memory kernel. 

Following conventions, the frequency $\Omega_0$ entering the Green's functions is not the \textit{bare} sub-system frequency, $\Omega_0=\omega_S$, but the \textit{renormalized} frequency $\Omega_\mathrm{ren}$,
\begin{equation}
\label{eq:renormalization}
\Omega_{\mathrm{ren}}^2
=
\Omega_0^2
+
\delta\Omega^2
\quad\quad
\delta\Omega^2
\simeq
\frac{2}{\pi}\int_{0}^{\infty} d\omega \,\frac{J(\omega)}{\omega},
\end{equation}
wherein $\delta\Omega^2$ denotes the (standard) bath-induced Lamb shift (see: \cite{breuer2002theory} pp. 530-535), a shift linear with the sub-system Hamiltonian regularly observed in experiment.

Thus, the evolution of sub-system observables is with respect to the renormalized sub-system Hamiltonian $H_{\mathrm{ren}}(t)$, where all contributions linear to the kinetic energy have been collected into the so-called physically observable frequency $\Omega_\mathrm{ren}(t)$, since the bare sub-system frequency has become indistinguishable from the bath-induced linear contributions in open-from, 
\begin{align}
H_{\mathrm{ren}}(t)
=
\frac{p^2}{2M}
+
\frac{1}{2}M\Omega_\text{ren}^2 x^2
-
\cancelto{V_c}{F_{\mathrm{eff}}(t)x} ,
\end{align} 
where the coherent forcing term $\propto F_{\mathrm{eff}}(t)x$ which would normally be appearing the expression above has now been absorbed into the bath spectral density in the form of a counter-potential $V_c$ which leads to \(-\mathcal{F}_E(t)\rho(t)\). 

The same spectral density that fixes the memory kernel fixes all counter-potentials.  In the stationary limit, this dependence is diagonal
in frequency, so that
\begin{equation}
    \delta \Omega^{2}
    =
    \frac{2}{\pi}
    \int_{0}^{\infty} d\omega\,
    \frac{J(\omega)}{\omega}.
\end{equation}

For a field-biased bath, the force correlations are not explicitly assumed to be stationary, and the
appropriate object is the two-time, or Wigner, spectral density
\begin{equation}
    J_{\rm eff}(\omega;T)
    =
    \int_{-\infty}^{\infty} d\tau\,
    e^{i\omega \tau}
    J_{\rm eff}
    \left(
    T+\frac{\tau}{2},
    T-\frac{\tau}{2}
    \right),
\end{equation}
with \(T=(t+t')/2\) and \(\tau=t-t'\).  Equivalently, the field-biased
renormalization entering the homogeneous Green's function is
$
    \Omega_{\rm ren}^{2}(T)
    =
    \Omega_{0}^{2}
    +
    \delta \Omega^{2}(T),
$
where
\begin{gather}
    \delta \Omega^{2}(T)
    =
    \frac{2}{\pi}
    \int_{0}^{\infty} d\omega\,
    \frac{J_{\rm eff}(\omega;T)}{\omega}.
\end{gather}
This expression makes explicit that the drive does not have to be inserted
only as a separate force term.  When the drive is treated quantum mechanically, it modifies the bath spectrum, and its effect can be carried by the two-time spectral density and appears as a
center-time-dependent Lamb shift.

Thus, since the master equation contains \(-\mathcal{F}_E(t)\rho(t)\), with the field contribution as
\begin{equation}
-\mathcal{F}_E^{(W)}(t)W
=
-\eta_E(t)\frac{\partial W}{\partial p}
+
\zeta_E(t)\frac{\partial W}{\partial x}.
\end{equation}
and the field-bias modification kernels,
\begin{gather}
\eta_E(t)=\int_0^t ds\,G_2(t-s)F_E(s),
\\
\zeta_E(t)=\int_0^t ds\,\dot{G}_2(t-s)F_E(s).
\end{gather} 
the term $F_{\mathrm{eff}}(t)x$ need not appear in $H_\mathrm{ren}(t)$. In this representation, the transformation effectively moves the dynamics into a frame co-evolving with the driven environment, such that the coherent displacement no longer appears as an external forcing term acting on the subsystem. Instead, the field-bias modifies the bath itself through a center-time-dependent renormalization.

\newpage  Importantly, the damping coefficient is solely determined by the inhomogeneous Green's function,
\begin{equation}
\Gamma(t)
=
\frac{1}{2}\frac{d}{dt}\ln \{G_2(t)\}.
\end{equation}

In contrast, the dissipation and diffusion coefficients depend explicitly on the full two-time correlators, 
\begin{align}
D_{xp}(t)
&=
\frac{\hbar}{M}
\int_{0}^{t} d\lambda \,
G_{1}(t,\lambda)\,
\nu(t,\lambda)
\\&\quad-
\frac{2\hbar}{M^{2}}
\int_{0}^{t} ds
\int_{s}^{t} d\tau
\int_{0}^{t} d\lambda \gamma(t,s)
\\
&\quad\quad\quad\quad\quad
\,
G_{1}(t,\lambda)\,
G_{2}(s,\tau)\,
\mu(\tau,\lambda)
\\
D_{pp}(t)
&=
\hbar
\int_{0}^{t} d\lambda \,
\partial_{t}G_{1}(t,\lambda)\,
\nu(t,\lambda)
\\
&\quad
-
\frac{2\hbar}{M}
\int_{0}^{t} ds
\int_{s}^{t} d\tau
\int_{0}^{t} d\lambda \, \gamma(t,s)\,
\\
&\quad\quad\quad\quad\quad
\dot G_{1}(t,\lambda)\,
G_{2}(s,\tau)\,
\mu(\tau,\lambda)
\end{align}
where $G_2(t)$ is the solution of the homogeneous equation, \textbf{Eq. (\ref{eq: LangevinBare})}. 
In the remaining sections, we'll analyze the long-time behavior of the kernels which generate these coefficients and determine the correction to the Markovian limit for monochromatic driving fields, comparing the resulting dynamics with those of the unbiased equilibrium. 

\section{Numerical Benchmarks}

\begin{table}
\label{tab:experimental_parameters}
\centering
\begin{ruledtabular}
\begin{tabular}{lcc}
Parameter & Range & Chosen \\
\hline
$\omega_p/2\pi$ 
& $4.0$--$8.0~\mathrm{GHz}$ & $5.0~\mathrm{GHz}$ \\
$|\mathrm{A}_\mathrm{RF}|^2$ 
& $0.01$--$10$ & $1.0$ \\
$T$ & $10$--$100~\mathrm{mK}$ & $20~\mathrm{mK}$ \\
\end{tabular}
\end{ruledtabular}
\caption{\textit{Drive frequencies.} Parameter set for a monochromatic drive. 
Drive frequencies in the $4$--$8~\mathrm{GHz}$ range match typical microwave control tones used in superconducting qubit experiments. 
Dimensionless drive amplitudes $|\alpha_p|^2$ spanning $0.01$--$10$ correspond to weak to strong coherent displacements achievable through standard microwave line driving \cite{koch2007transmon,reagor2016quantum}. 
The temperature range $10$--$100~\mathrm{mK}$ is consistent with dilution refrigerator operating conditions.}
\label{tab:drive_parameters}
\centering
\end{table}

For a monochromatic field bias, the general kernel expressions derived above acquire an explicitly oscillatory two-time structure and therefore provide a direct setting in which to test the field-biased HPZ construction. 

We evaluate the exact expressions for the driven force correlator, noise kernel, and reduced master-equation coefficients for the harmonic potential, and compare them with the direct numerical calculations of the same quantities. The GLE used for the simulations is 
\begin{equation}
    \ddot{x}(t)
    +
    \Omega_{0}^{2}x(t)
    +
    \int_{0}^{t}ds\,
    \gamma(t,s)\dot{x}(s)
    =
    \frac{1}{m}F_{\mathrm{ext}}(t).
\end{equation}

\begin{figure}
    \centering
    \begin{overpic}[width=\linewidth]{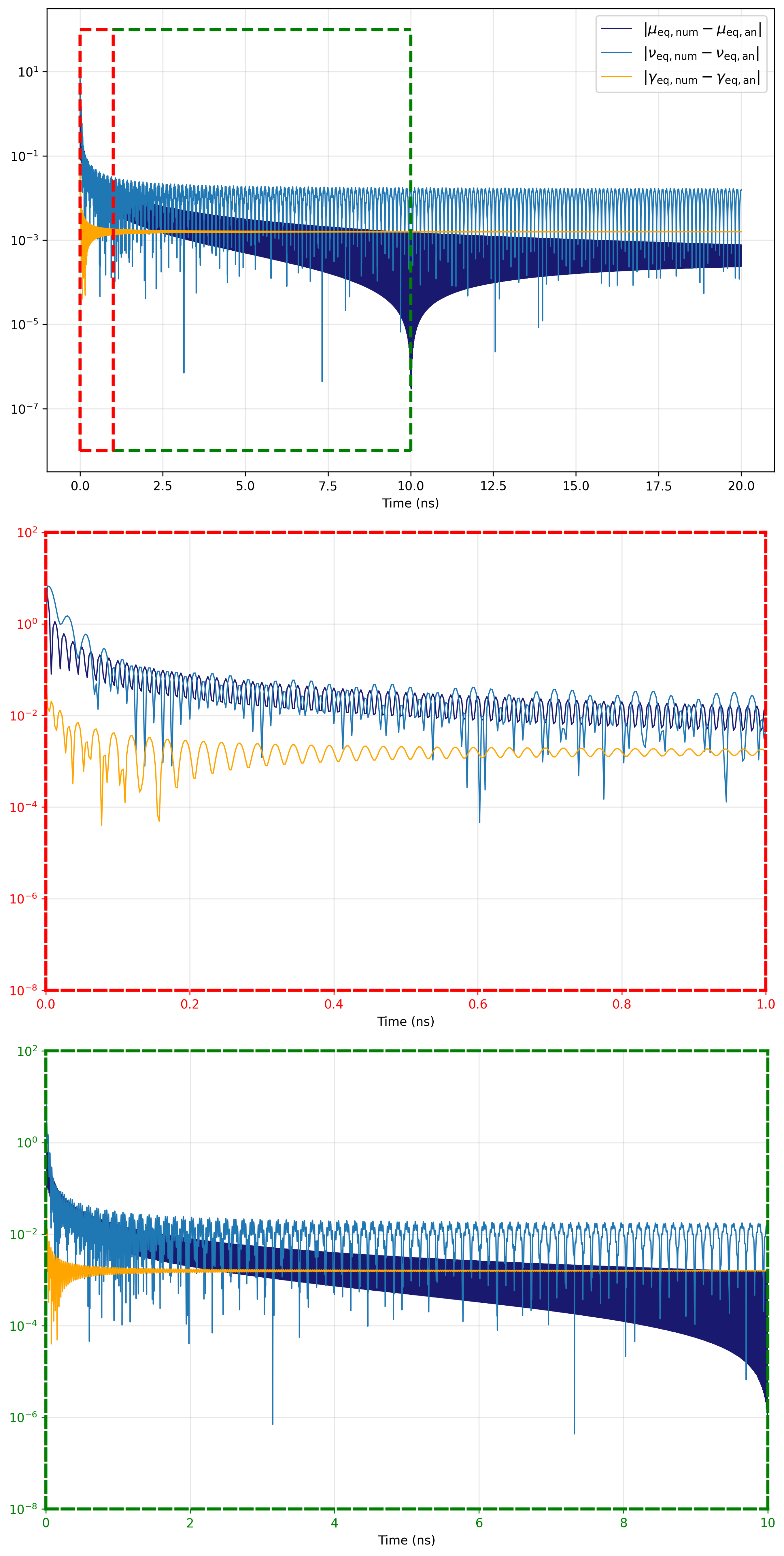}
        \put(45,73){\bfseries (a)}
        \put(45,62){\bfseries (b)}
        \put(45,29){\bfseries (c)}
    \end{overpic}
        \caption{\textit{Equilibrium numerical performance.} Comparison between the numerical and analytical kernel solutions used throughout the generalized Langevin equation simulations. \textbf{(a)} full-time evolution of the absolute deviations between the numerical and analytical equilibrium kernels, including the memory kernel $\gamma(t)$, dissipation kernel $\eta(t)$, and noise kernel $\nu(t)$. \textbf{(b)} short-time transient behavior highlighted by the red dashed region, illustrating the initial non-Markovian relaxation and rapid oscillatory structure. \textbf{(c)} intermediate- to long-time dynamics corresponding to the green dashed region, where the analytical and numerical solutions remain in close agreement over the full simulation interval. All quantities are plotted on logarithmic scales to emphasize both transient and asymptotic deviations. Notably, the code resolves to a minimum absolute distance from the equilibrium.}
    \label{fig: ERR}
\end{figure}

The nonlocal damping is determined by the memory kernel \(\gamma(t,s)\). In the stationary equilibrium limit, this kernel depends only on the time difference \(\tau=t-s\), 
\begin{equation}
    \gamma(t,s)=\gamma(\tau),
    \qquad
    \tau=t-s\geq 0.
\end{equation}

The equilibrium spectral density is taken to be of the Lorentz--Drude form as denoted in \textbf{Eq. (\ref{eq: Lorentz})} where \(f\) is measured in GHz to reflect physical circuit quantities, \(\lambda\) is the cutoff frequency, and \(\gamma\) fixes the overall damping scale. The memory kernel is computed from,
\begin{equation}
    \gamma_{\mathrm{eq}}(\tau)
    =
    \frac{2}{\pi}
    \int_{0}^{f_{\max}}
    df\,
    \frac{J_{\mathrm{eq}}(f)}{f}
    \cos(2\pi f\tau).
\end{equation}

Numerically, this is evaluated on a uniform frequency grid \(f_{j}\) by trapezoidal quadrature,
\begin{equation}
    \gamma_{\mathrm{eq}}(\tau_{i})
    \approx
    \frac{2}{\pi}
    \sum_{j}
    w_{j}
    \frac{J_{\mathrm{eq}}(f_{j})}{f_{j}}
    \cos(2\pi f_{j}\tau_{i}),
\end{equation}
with the \(f=0\) point removed from the quotient \(J(f)/f\) and replaced by its finite limiting value.

The corresponding dissipation kernel is computed directly from the spectral density as
\begin{equation}
    \nu_{\mathrm{eq}}(\tau)
    =
    \int_{0}^{f_{\max}}
    df\,
    J_{\mathrm{eq}}(f)
    \sin(2\pi f\tau).
\end{equation}

For consistency, we also evaluate it from the memory kernel through
\begin{equation}
    \nu_{\mathrm{eq}}(\tau)
    =
    -\frac{1}{4}
    \frac{d}{d\tau}
    \gamma_{\mathrm{eq}}(\tau),
    \label{eq: identity}
\end{equation}
where the derivative is computed by a centered finite-difference rule on the time grid. The symmetrized noise kernel is
\begin{equation}
    \mu_{\mathrm{eq}}(\tau)
    =
    \int_{0}^{f_{\max}}
    df\,\nu_T 
    J_{\mathrm{eq}}(f)
    \cos(2\pi f\tau).
\end{equation}
where the thermal factor is 
\begin{equation}
    \nu_T =     \coth\!\left(
    \frac{hf10^{9}}{2k_{\mathrm{B}}T}
    \right)
\end{equation}
At \(f=0\), the product \(J(f)\coth[hf10^{9}/(2k_{\mathrm{B}}T)]\),
\begin{equation}
    \lim_{f\rightarrow 0}
    J(f)
    \coth\!\left(
    \frac{hf10^{9}}{2k_{\mathrm{B}}T}
    \right)
    =
    \left[
    \lim_{f\rightarrow 0}
    \frac{J(f)}{f}
    \right]
    \frac{2k_{\mathrm{B}}T}{h10^{9}}.
\end{equation}
 is evaluated by its central limit. Conveniently, the operating temperature of a dillution refrigerator, which is our chosen environmental equilibrium temperature admits a particularly simple form since it is nearly the zero-temperature kernel. 

For the classical field-bias, the drive is treated as a deterministic coherent displacement,
\begin{equation}
    F_{\mathrm{drv}}(t)
    =
    F_{0}\cos(2\pi f_{d}t+\phi_{d}).
\end{equation}

It is not inserted into the connected memory kernel. Therefore
$
    \gamma_{\mathrm{cl}}(\tau)
    =
    \gamma_{\mathrm{eq}}(\tau),
    \nu_{\mathrm{cl}}(\tau)
    =
    \nu_{\mathrm{eq}}(\tau).
$.  The uncentered force AoC receives the coherent contribution
\begin{equation}
    C_{\mathrm{drv}}(\tau)
    =
    A_{\mathrm{drv}}F_{0}^{2}
    \cos(2\pi f_{d}\tau),
\end{equation}
where \(A_{\mathrm{drv}}\) is the numerical phase-averaging pre-factor used in the simulation. Hence
\begin{equation}
    \mu_{\mathrm{cl}}(\tau)
    =
    \mu_{\mathrm{eq}}(\tau)
    +
    C_{\mathrm{drv}}(\tau).
\end{equation}

The connected spectral density is decomposed as in
$
    J_{\mathrm{qm}}(f)
    =
    J_{\mathrm{eq}}(f)
    +
    J_{\mathrm{coh}}(f)
    +
    J_{\mathrm{inc}}(f).
$
The coherent part is represented numerically through a narrow normalized Gaussian,
\begin{equation}
    J_{\mathrm{coh}}(f)
    =
    A_{\mathrm{coh}}
    \frac{
    \exp[-(f-f_{\mathrm{coh}})^{2}/(2\mu_{\mathrm{coh}}^{2})]
    }{
    \sqrt{2\pi}\mu_{\mathrm{coh}}
    },
\end{equation}
while the incoherent pump contribution term is modeled as a de-tuned Lorentzian,
\begin{equation}
    J_{\mathrm{inc}}(f)
    =
    A_{\mathrm{inc}}
    \frac{
    \Lambda_{\mathrm{inc}}^{2}
    }{
    (f-f_{\mathrm{coh}})^{2}+\Lambda_{\mathrm{inc}}^{2}
    }.
\end{equation}
Only the connected spectrum $J_\mathrm{con}(f)$ is used to build the inhomogeneous memory and dissipation kernels.

The stationary kernels are explicitly two-time causal kernels in the interpolation scheme:
\begin{equation}
    K(t,s)
    =
    H(t-s)K(t-s),
\end{equation}
where \(K\in\{\gamma,\nu,\mu\}\), meaning
\begin{equation}
    K(t_{i},s)
    =
    \begin{cases}
    \mathrm{Interp}\!\left[K(\tau)\right]_{\tau=t_{i}-s},
    &
    t_{i}\geq s,\\
    0,
    &
    t_{i}<s.
    \end{cases}
\end{equation}
The memory integral appearing in the GLE is then evaluated as
\begin{equation}
    \int_{0}^{t_{i}}ds\,
    \gamma(t_{i},s)\dot{x}(s)
    \simeq
    \sum_{j\leq i}
    w_{j}
    \gamma(t_{i},t_{j})\dot{x}(t_{j}).
\end{equation}

\begin{figure}
    \centering

    \begin{overpic}[width=\linewidth]{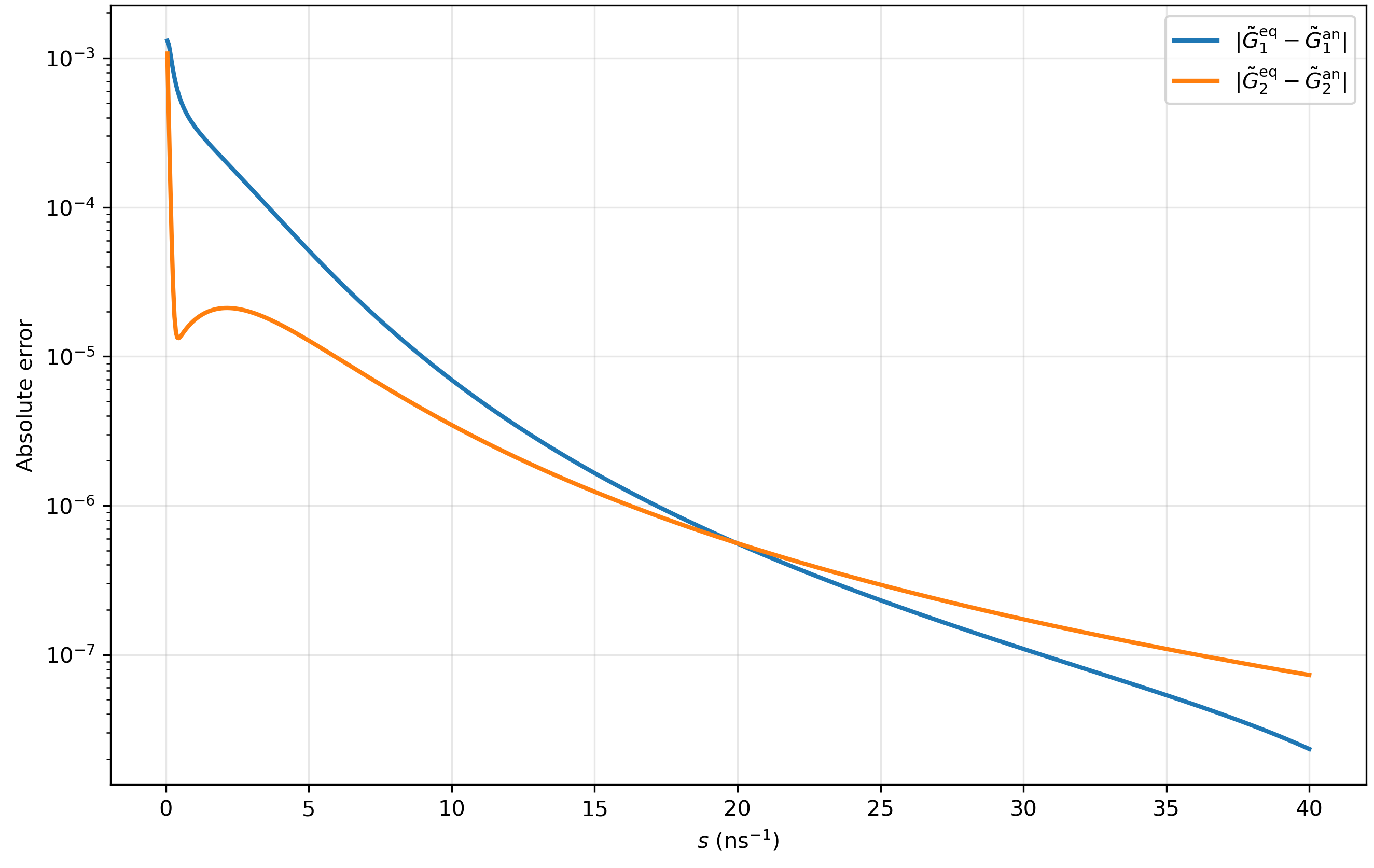}
        \put(17.5,55){\bfseries (a)}
    \end{overpic}

    \vspace{0.5em}

    \begin{overpic}[width=\linewidth]{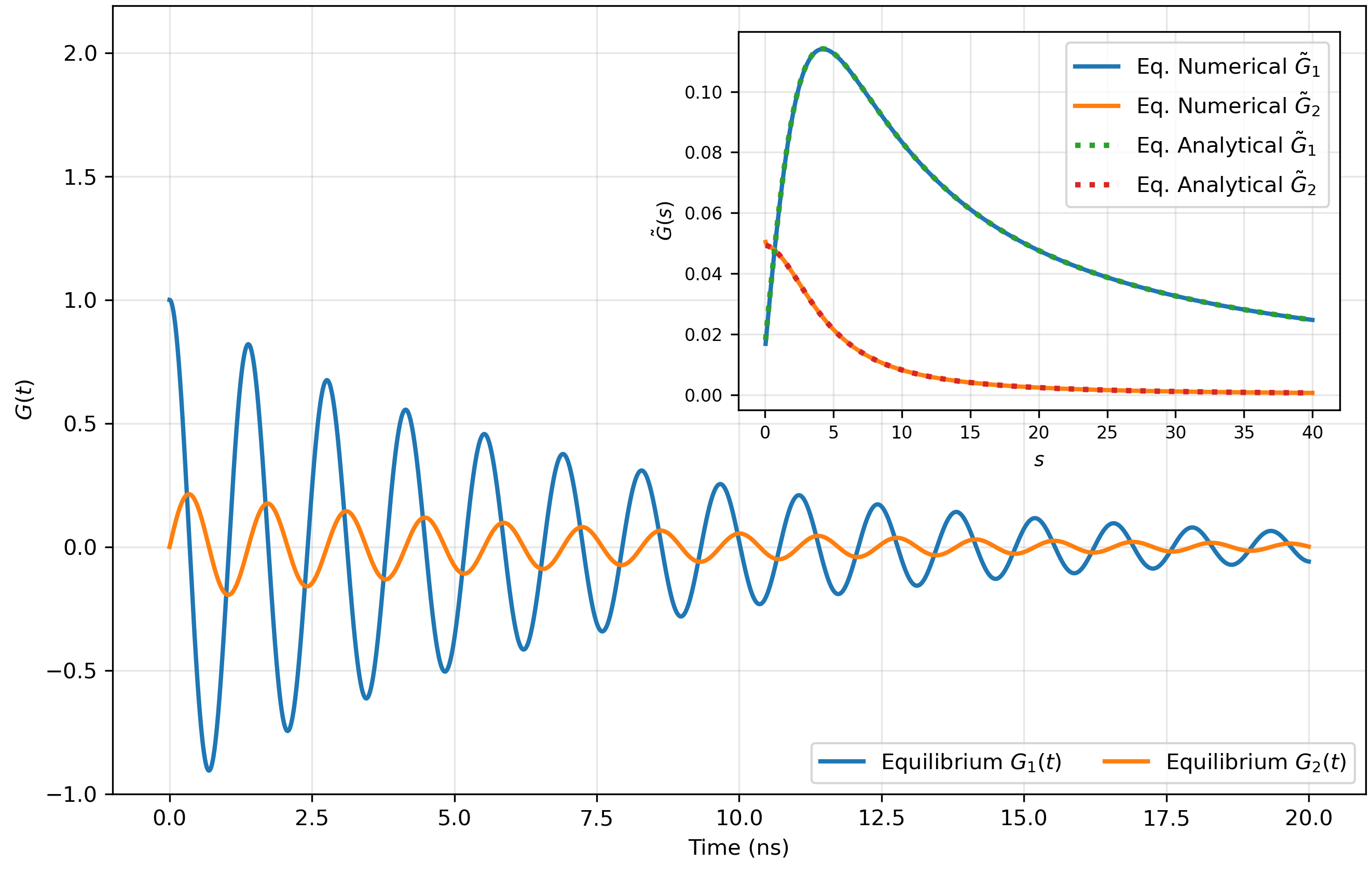}
        \put(12,55){\bfseries (b)}
    \end{overpic}

    \caption{\textit{Green's functions in Laplace space.} Comparison of the numerical and analytical Green functions: \textbf{(a)} shows Laplace-space absolute errors, while \textbf{(b)} shows time-domain homogeneous Green functions with agreement in the inset, demonstrating strong numerical consistency in Laplace space.}
    \label{fig: GreensLaplace}
\end{figure}

Setting $\Lambda_\mathrm{inc}=$~$F_0=0$ thus recovers the undriven, equilibrium kernels which are utilized to obtain the overall accuracy of the solver reported in \textbf{Figure \ref{fig: ERR}}, confirming reliable numerical accuracy.  

We now utilize the code to study two examples. In the first example, the coherent spectrum ($J_\mathrm{coh}(\omega)$) is excluded from the connected inhomogeneous kernel to avoid double-counting the deterministic displacement. In the second example, the full spectral density is utilized and $F_0=0$. The primary objective of these examples is to explicitly compare how equilibrium, classical-drive, and quantum-drive treatments modify the reduced subsystem dynamics at the level of the memory, dissipation and nosie kernel generators of the coefficients of the master equation. In particular, the simulations are designed to isolate which features of the dynamics originate from coherent displacement, which arise from modifications to the environmental spectral structure, and which correspond to genuine nonstationary effects induced by field bias.


\subsection{Analytical validation of the Green's functions}

All benchmark calculations are performed within the same underlying parameters, allowing direct comparison between the different driving prescriptions. The equilibrium case serves as the stationary reference solution, while the classical and quantum treatments introduce progressively richer forms of environmental dressing through coherent forcing and structured spectral modifications, requiring a number of benchmarks. The Green's functions \(G_{1}(t)\) and \(G_{2}(t)\) are obtained by solving the homogeneous equations with
\begin{gather}
    G_{1}(0)=1,
    \qquad
    \dot{G}_{1}(0)=0,
\\
    G_{2}(0)=0,
    \qquad
    \dot{G}_{2}(0)=\frac{1}{m_{S}}.
\end{gather}
The solution is then reconstructed as
\begin{align}
    x(t)
    &=
    G_{1}(t)x(0)
    +
    G_{2}(t)p(0)
    \\&\quad\quad+
    \int_{0}^{t}ds\,
    G_{2}(t-s)F_{\mathrm{ext}}(s).
\end{align}
The analytical solutions follow directly from solving the GLE in Laplace space \cite{breuer2002theory}. Taking the Laplace transform converts the GLE into an algebraic equation,
\begin{equation}
m z^2 \tilde{x}(z) + z \tilde{\gamma}(z)\tilde{x}(z) + m\omega^2 \tilde{x}(z) = \tilde{F}(z),
\end{equation}
so that the response function is
\begin{equation}
\tilde{G}(z) = \frac{1}{m z^2 + z \tilde{\gamma}(z) + m\omega^2}.
\end{equation}
The functions \(G_1(t)\) and \(G_2(t)\) are then identified as the inverse Laplace transforms associated with this propagator, fixed by the initial conditions of the homogeneous equation. In Laplace space, they are of the simple analytical form (see: \textbf{Figure \ref{fig: GreensLaplace}.})
\begin{gather}
\label{eq: G1}
\tilde G_{2}(\Omega)
=
\frac{1}{m_{S}\!\left(\Omega^{2}
+
\Omega\,\tilde\gamma(\Omega)
+
\Omega_{0}^{2}\right)},
\\
\label{eq: G2}
\tilde G_{1}(\Omega)
=
\frac{\Omega+\tilde\gamma(\Omega)}
{\Omega^{2}
+
\Omega\,\tilde\gamma(\Omega)
+
\Omega_{0}^{2}},
\end{gather}
where we have expressed the noise kernel in the susceptibility (frequency) domain, 
\begin{equation}
\tilde\gamma(\Omega)
=
\int_{0}^{\infty} dt\, e^{-\Omega t}\,\gamma(t).
\end{equation} 

For the homogeneous part, the analytical solution can be written as
\begin{equation}
x^{(h)}(t) = G_1(t)\,x(0) + G_2(t)\,p(0).
\end{equation}

For the local undamped harmonic oscillator with frequency $\Omega_S$,
\begin{equation}
G_1(t) = \cos(\Omega_S t),
\qquad
G_2(t) = \frac{\sin(\Omega_S t)}{m_S \Omega_S}.
\end{equation}

Therefore,
\begin{equation}
{
x^{(h)}(t)
=
x(0)\cos(\Omega_S t)
+
\frac{p(0)}{m_S \Omega_S}\sin(\Omega_S t)
}
\end{equation}
This form of the homogeneous solution holds for a linear harmonic oscillator with constant frequency, where the dynamics are time-local. Together with \textbf{Eq. (\ref{eq: G1})-(\ref{eq: G2})}, we can benchmark the numerical accuracy of the Green's function reconstruction reported in \textbf{Figure \ref{fig: GreensErrors}.} 

\subsection{Representing coherent displacements}
\begin{figure}
    \centering

    \begin{overpic}[width=\linewidth]{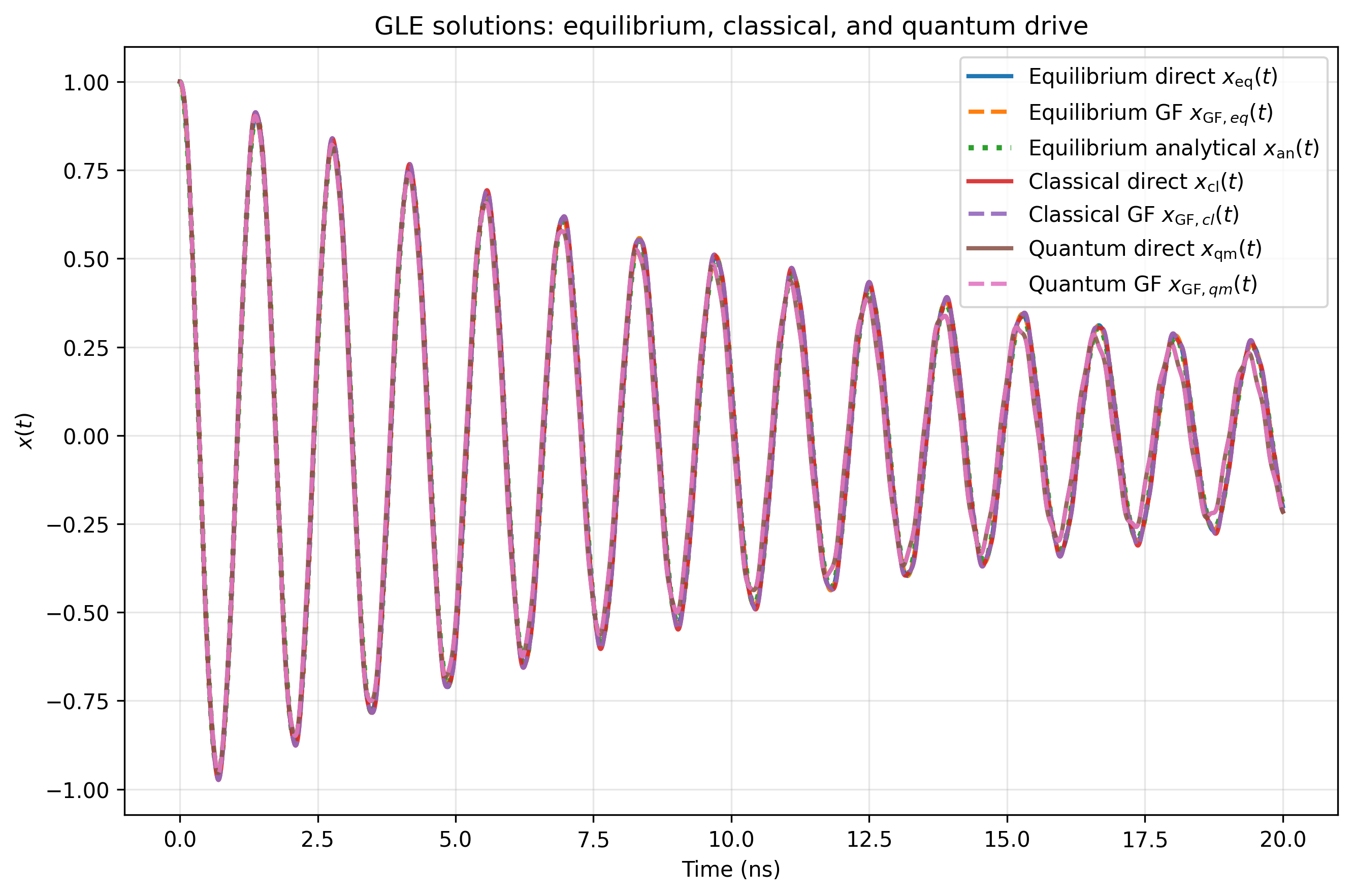}
        \put(85,15){\bfseries (a)}
    \end{overpic}

    \vspace{0.5em}

    \begin{overpic}[width=\linewidth]{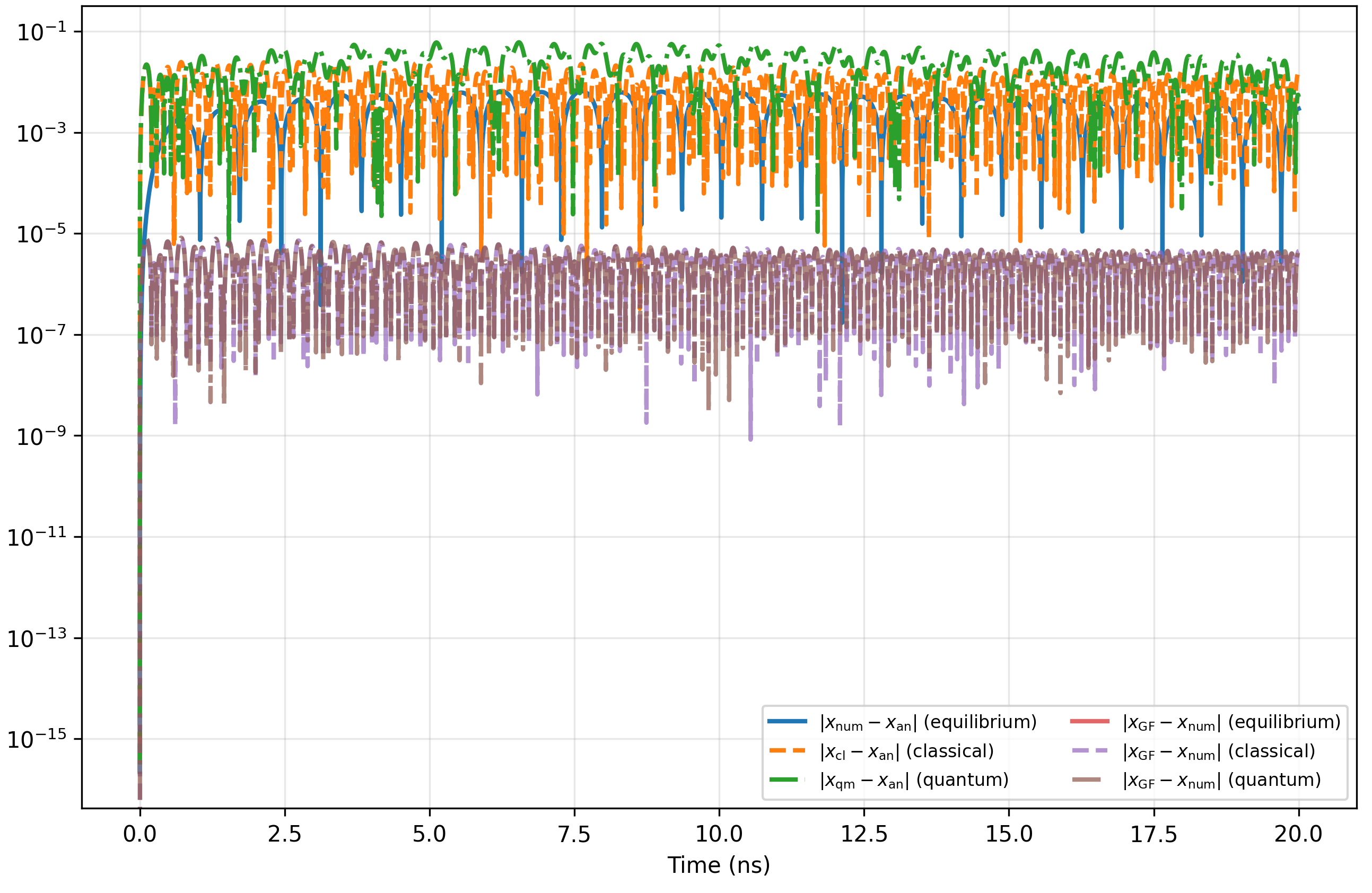}
        \put(85,30){\bfseries (b)}
    \end{overpic}

    \caption{\textit{Green's functions errors.} The comparison of numerical and the analytical Green functions: plot \textbf{(a)} shows the absolute errors in the Green's function reconstruction, while plot \textbf{(b)} shows time-domain homogeneous Green functions with inset agreement, demonstrating strong numerical–analytical consistency. After $t=10$ The reconstruction begins to gradually come out of phase with the analytical solution as a natural consequence of the truncated interpolation scheme, marking the largest source of numerical error.}
    \label{fig: GreensErrors}
    \label{fig: GreensErrors}
\end{figure}
\begin{figure}
    \centering

    \begin{overpic}[width=\linewidth]{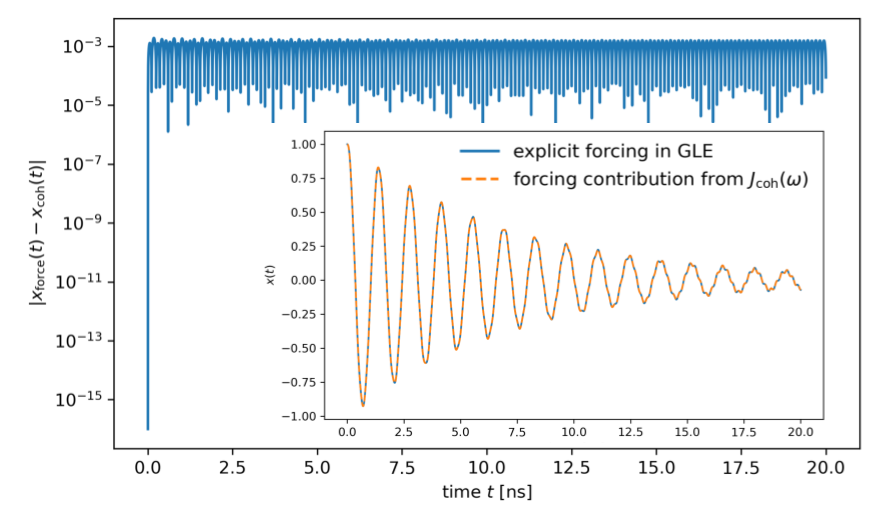}
        \put(25,35){\bfseries (a)}
    \end{overpic}

    \vspace{0.5em}

    \begin{overpic}[width=\linewidth]{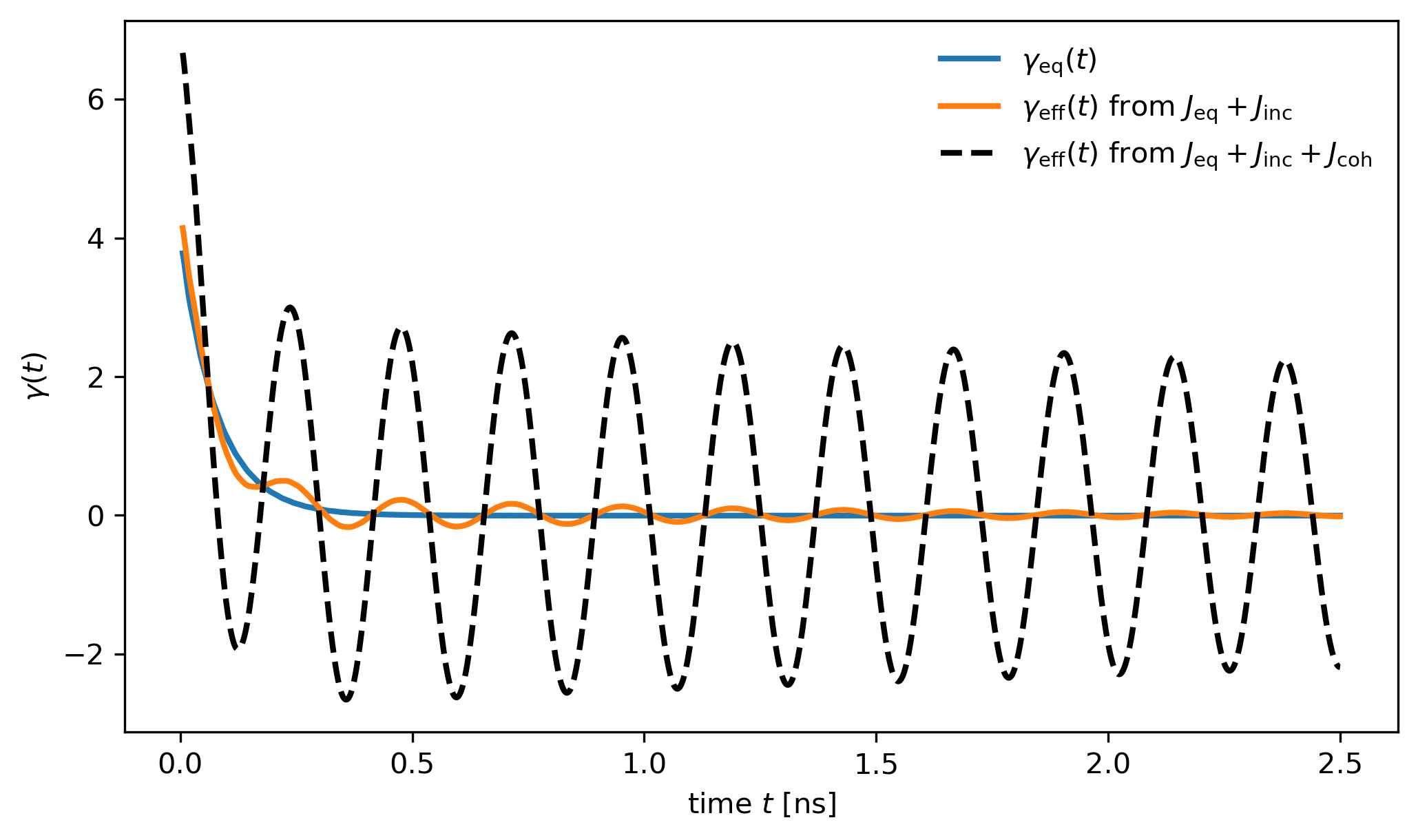}
        \put(20,45){\bfseries (b)}
    \end{overpic}

    \caption{\textit{Coherent focing.} Coherent forcing equivalence in the GLE, showing \textbf{(a)} absolute difference between the oscillator trajectories obtained from an explicit periodic forcing term in the GLE and from the coherent spectral contribution \(J_{\rm coh}(\omega)\), demonstrating numerical equivalence to near machine precision. The inset shows the corresponding trajectories directly, confirming that the coherent spectral density reproduces the same driven displacement dynamics as the explicit forcing term. \textbf{(b)} comparison of the equilibrium memory kernel \(\gamma_{\rm eq}(t)\), the kernel reconstructed from \(J_{\rm eq}+J_{\rm inc}\), and the kernel reconstructed from the full spectrum \(J_{\rm eq}+J_{\rm inc}+J_{\rm coh}\). The coherent contribution produces persistent oscillatory structure associated with the monochromatic drive, while the dissipative background remains governed by the equilibrium and incoherent spectral components. }
    \label{fig: coherentforcing}
\end{figure}

%




To demonstrate that the coherent displacement may be generated through the coherent contribution to the force autocorrelation rather than through an explicit forcing term, we compare two equivalent implementations of the driven generalized Langevin equation. The reference implementation evolves
\begin{equation}
    \ddot{x}(t)
    +
    \Omega_{0}^{2}x(t)
    +
    \int_{0}^{t}ds\,
    \gamma(t-s)\dot{x}(s)
    =
    \frac{1}{m}F_{\mathrm{drv}}(t),
\end{equation}
with
\begin{equation}
    F_{\mathrm{drv}}(t)
    =
    F_{0}\cos(2\pi f_{d}t+\phi_{d}).
\end{equation}
In this form, the coherent tone is treated as a deterministic force acting directly on the oscillator.

The modified code removes this explicit forcing term and instead places the coherent contribution into the uncentered force spectrum. The equation of motion is then evolved as
\begin{equation}
    \ddot{x}(t)
    +
    \Omega_{0}^{2}x(t)
    +
    \int_{0}^{t}ds\,
    \gamma_{\mathrm{conn}}(t-s)\dot{x}(s)
    =
    0,
\end{equation}
while the coherent displacement is reconstructed from the coherent force autocorrelation. The total force spectrum is decomposed as
\begin{equation}
    J_{\mathrm{tot}}(f)
    =
    J_{\mathrm{eq}}(f)
    +
    J_{\mathrm{inc}}(f)
    +
    J_{\mathrm{coh}}(f),
\end{equation}
but only the connected part
\begin{equation}
    J_{\mathrm{conn}}(f)
    =
    J_{\mathrm{eq}}(f)
    +
    J_{\mathrm{inc}}(f)
\end{equation}
is used to construct the memory and dissipation kernels, as shown below,
\begin{equation}
    \gamma_{\mathrm{conn}}(\tau)
    =
    \frac{2}{\pi}
    \int_{0}^{f_{\max}}
    df\,
    \frac{J_{\mathrm{conn}}(f)}{f}
    \cos(2\pi f\tau),
\end{equation}
\begin{equation}
    \nu_{\mathrm{conn}}(\tau)
    =
    \int_{0}^{f_{\max}}
    df\,
    J_{\mathrm{conn}}(f)
    \sin(2\pi f\tau).
\end{equation}
The coherent peak is excluded from these connected kernels because it represents a mean field, not a fluctuating dissipative bath. Including it in \(\gamma\) would incorrectly allow the coherent displacement to renormalize the homogeneous damping.

The coherent part is instead inserted into the uncentered force autocorrelation,
\begin{equation}
    C_{\mathrm{coh}}(t,t')
    =
    F_{\mathrm{coh}}(t)F_{\mathrm{coh}}(t'),
\end{equation}
with
\begin{equation}
    F_{\mathrm{coh}}(t)
    =
    F_{0}\cos(2\pi f_{d}t+\phi_{d}).
\end{equation}
Equivalently, after phase averaging,
\begin{equation}
    C_{\mathrm{coh}}(\tau)
    =
    \frac{F_{0}^{2}}{2}
    \cos(2\pi f_{d}\tau).
\end{equation}

In frequency space, this corresponds to the coherent spectral contribution
\begin{equation}
    S_{\mathrm{coh}}(f)
    =
    \frac{F_{0}^{2}}{4}
    \left[
    \delta(f-f_{d})
    +
    \delta(f+f_{d})
    \right],
\end{equation}
which is represented numerically by a narrow Gaussian peak,
\begin{equation}
    J_{\mathrm{coh}}(f)
    =
    A_{\mathrm{coh}}
    \frac{
    \exp[-(f-f_{d})^{2}/(2\mu_{\mathrm{coh}}^{2})]
    }{
    \sqrt{2\pi}\mu_{\mathrm{coh}}
    }.
\end{equation}

The numerical evidence consists of two simulations. In the first simulation, the coherent tone appears as an explicit force:
\begin{equation}
    x_{\mathrm{force}}(t)
    =
    x_{\mathrm{hom}}(t)
    +
    \int_{0}^{t}ds\,
    G_{2}(t-s)F_{\mathrm{drv}}(s).
\end{equation}
In the second simulation, the forcing function is set to zero, but the same coherent contribution is retained in the uncentered force sector and converted into the equivalent displacement response:
\begin{equation}
    x_{\mathrm{coh-spec}}(t)
    =
    x_{\mathrm{hom}}(t)
    +
    \int_{0}^{t}ds\,
    G_{2}(t-s)F_{\mathrm{coh}}(s).
\end{equation}
Since both constructions use the same Greens function, they must agree whenever the coherent spectral peak is normalized so that it corresponds to the same deterministic force amplitude \(F_{0}\). The code therefore verifies
\begin{equation}
    \Delta x(t)
    =
    x_{\mathrm{force}}(t)
    -
    x_{\mathrm{coh-spec}}(t)
    \approx
    0,
\end{equation}
and reports the maximum absolute error
\begin{equation}
    \epsilon_{\max}
    =
    \max_{t}
    \left|
    x_{\mathrm{force}}(t)
    -
    x_{\mathrm{coh-spec}}(t)
    \right|.
\end{equation}

\begin{figure*}[t]

    \begin{overpic}[width=\linewidth]{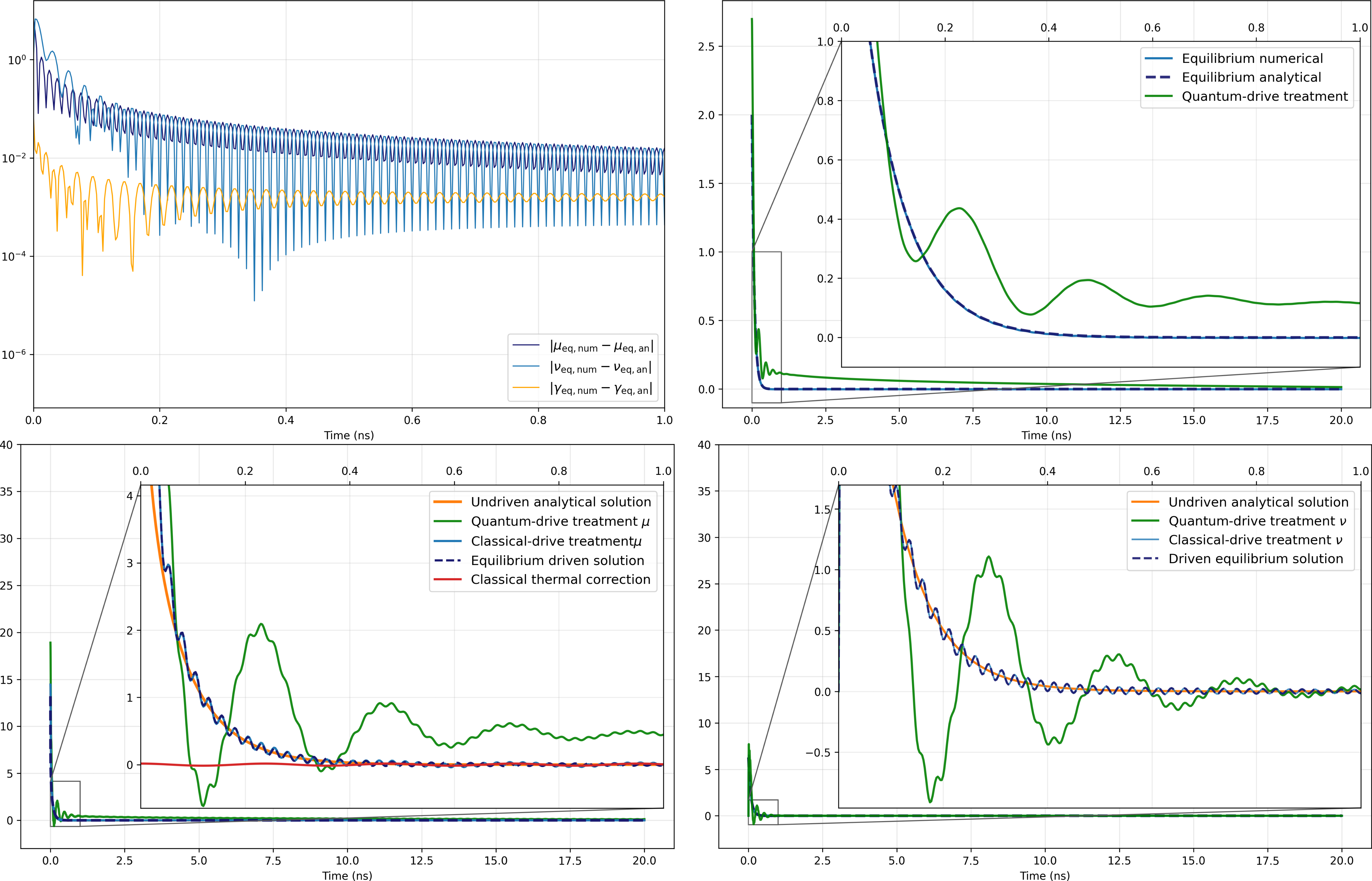}
        \put(5,60){\bfseries (a)}
        \put(56,60){\bfseries (b)}
        \put(2,28){\bfseries (c)}
        \put(54,28){\bfseries (d)}
    \end{overpic}

\centering
\caption{\textit{Kernels under periodic forcing.}
Comparison between equilibrium, classical-drive, and quantum-drive treatments of the reconstructed GLE kernels and their corresponding dynamical corrections under the classical and quantum mechanical treatments of the drive. \textbf{(a)} Absolute deviations between the numerical and analytical equilibrium kernels, showing agreement over the full simulation interval for the memory kernel $\gamma(t)$, dissipation kernel $\eta(t)$, and noise kernel $\nu(t)$. The largest discrepancies occur at early times where the rapidly oscillating non-Markovian transients are most pronounced. \textbf{(b)} Memory kernel comparison. The equilibrium numerical and analytical solutions remain nearly indistinguishable, while the quantum-drive treatment develops long-lived oscillatory corrections arising from the explicitly nonstationary bath spectral density. \textbf{(c)} Dissipation kernel comparison. The classical-drive treatment overlaps the equilibrium solution, demonstrating that coherent displacement acts primarily as a forcing correction and does not significantly modify the homogeneous dissipative structure. In contrast, the quantum-drive treatment produces a persistent nonequilibrium modulation. The classical thermal correction remains comparatively small throughout the evolution. \textbf{(d)} Noise kernel comparison. The equilibrium and classical-drive solutions again coincide, whereas the quantum-drive treatment exhibits slowly decaying oscillations induced by the drive-biased bath correlations. Insets in panels \textbf{(b)}–\textbf{(d)} magnify the short-time transient regime, where the deviations from the equilibrium stationary limit are most visible.
}
\label{fig: panel1}
\end{figure*}

\begin{figure*}[t]

    \begin{overpic}[width=\linewidth]{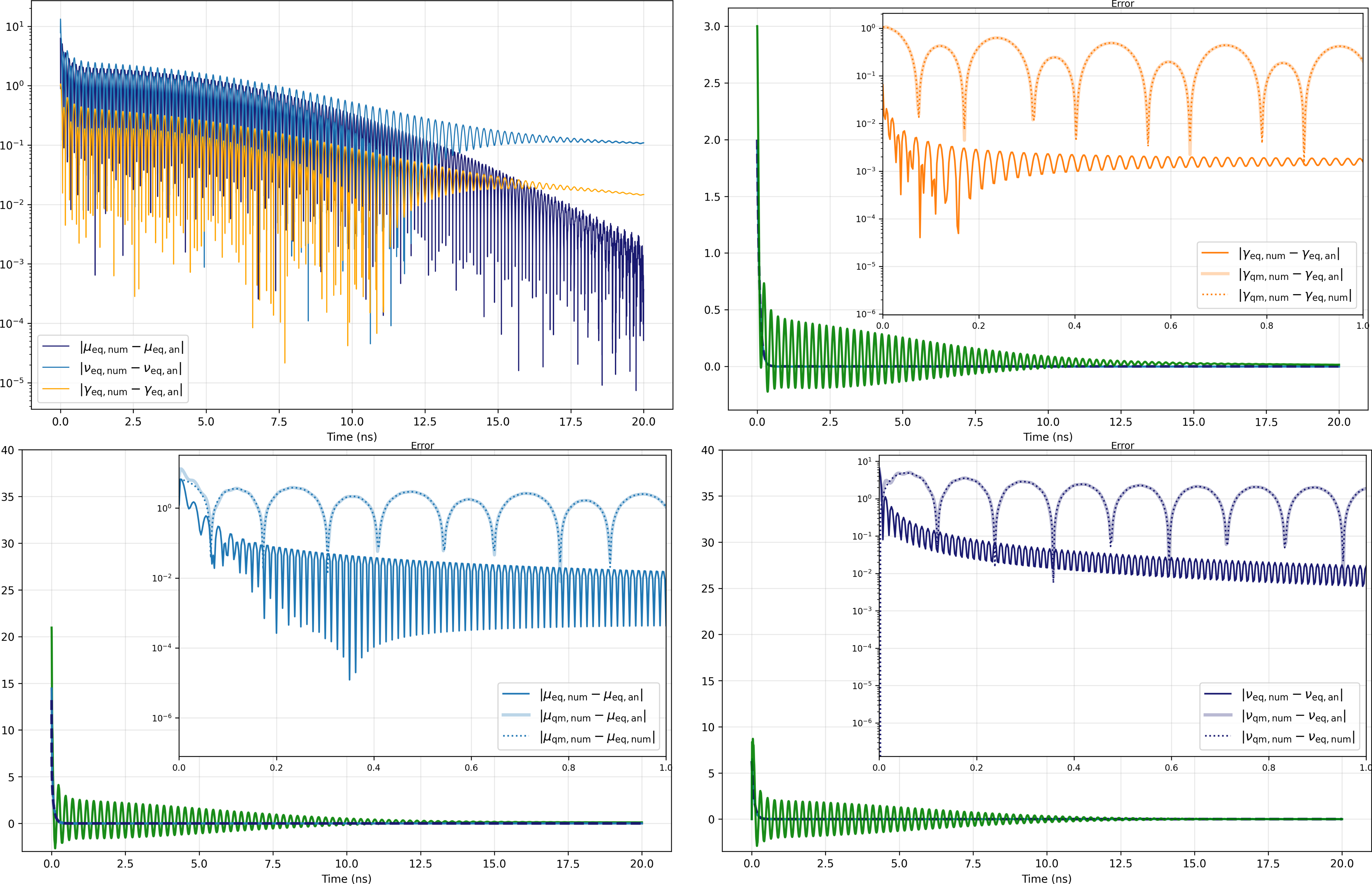}
        \put(6,62){\bfseries (a)}
        \put(56,60){\bfseries (b)}
        \put(3,28){\bfseries (c)}
        \put(54,28){\bfseries (d)}
    \end{overpic}
\centering
\caption{\textit{Periodic forcing through the spectral density.} Comparison of equilibrium, rotating-frame, and quantum-drive kernel dynamics illustrating the equivalence between the equilibrium driven and undriven descriptions under a reference-frame transformation. \textbf{(a)} Absolute deviations between the numerical and analytical equilibrium kernels in the laboratory frame, showing the oscillatory structure inherited from the explicit coherent drive. \textbf{(b)} Memory kernel comparison after transforming into the rotating frame of the coherent displacement. The equilibrium driven and undriven solutions become indistinguishable, demonstrating that the coherent contribution corresponds primarily to a change of reference frame rather than a modification of the underlying homogeneous dissipative dynamics. The remaining quantum-drive contribution exhibits persistent oscillatory corrections arising from the explicitly nonstationary bath spectrum. \textbf{(c)} Dissipation kernel comparison in the rotating frame. The collapse of the equilibrium driven and undriven curves confirms that the coherent displacement can be absorbed into the bath representation without altering the stationary equilibrium kernel itself. The inset highlights the residual numerical error between the transformed equilibrium solutions, which remains several orders of magnitude smaller than the kernel amplitudes. \textbf{(d)} Noise kernel comparison under the same transformation. The driven equilibrium and undriven equilibrium solutions again coincide after the rotating-frame mapping, while the quantum-drive treatment retains slowly decaying nonequilibrium oscillations induced by the field-biased bath correlations. Insets magnify the short-time transient regime where the equivalence between the transformed equilibrium solutions and the persistence of the nonstationary quantum-drive corrections are most clearly visible.
}
\label{fig: panel2}
\end{figure*}

As illustrated in \textbf{Figure \ref{fig: coherentforcing}.}, this modification proves that the coherent displacement does not need to enter as an explicit forcing term. It may instead be represented through the coherent contribution to the uncentered force spectrum, provided that the coherent peak is excluded from the connected memory and dissipation kernels. This is exactly the distinction already implemented in the benchmark code: \(J_{\mathrm{eq}}+J_{\mathrm{inc}}\) builds the connected kernels, while the coherent contribution is retained only as a displacement.

\subsection{Numerical evaluation of the kernels}\label{sec: subharmonic_perturbations}

All of the reconstructed kernels were benchmarked against their exact analytical expressions derived from the spectral density. This provides a direct consistency check of both the quadrature routines and the implementation of the generalized Langevin equation. Starting from the definition of the spectral density, \(J(f)\). 

For the Lorentz--Drude spectral density used throughout the simulations,
\begin{equation}
    J(f)
    =
    \frac{2\gamma\lambda^{2} f}{f^{2}+\lambda^{2}},
\end{equation}
all kernels admit well-known \cite{caldeira1983path, breuer2002theory} closed-forms. The memory and its connected kernels can be obtained using standard identities from the theorem of residues
\begin{align}
\gamma(\tau)
&=
2\gamma\lambda e^{-2\pi\lambda \tau},
\\
\mu(\tau)
&=
-\frac{d\gamma(\tau)}{d\tau}
=
4\pi\gamma\lambda^{2}e^{-2\pi\lambda \tau},
\\
\nu(\tau)
&=
\frac{2\gamma\lambda^{2}}{\beta\hbar}
\Big[
\cot\!\left(\pi\beta\hbar\lambda\right)
e^{-2\pi\lambda \tau}
\\&\quad\quad\quad+
2
\sum_{n=1}^{\infty}
\frac{\nu_{n}e^{-\nu_{n}\tau}}
{\nu_{n}^{2}-(2\pi\lambda)^{2}}
(2\pi\lambda)
\Big],
\end{align}
where the Matsubara frequencies are given by $\nu_{n}$.
The numerical kernels \(\gamma_{\mathrm{num}}(\tau)\), \(\nu_{\mathrm{num}}(\tau)\), and \(\mu_{\mathrm{num}}(\tau)\) were constructed using discrete quadrature over a finite frequency window \([0,f_{\max}]\). The numerical validation was performed by computing the point-wise deviations
\begin{gather}
    \Delta_{\epsilon}(\tau)
    =
    \gamma_{\mathrm{num}}(\tau)
    -
    \epsilon(\tau),
\end{gather}
for $\epsilon=\{\gamma, \mu, \mu\}$ and verifying that
\begin{equation}
    \max_{\tau}
    \left|
    \Delta_{\gamma}(\tau)
    \right|,
    \;
    \max_{\tau}
    \left|
    \Delta_{\nu}(\tau)
    \right|,
    \;
    \max_{\tau}
    \left|
    \Delta_{\mu}(\tau)
    \right|
    \sim 10^{-12},
\end{equation}
is consistent up to numerical precision. An additional internal consistency check follows from the identity in \textbf{Eq. (\ref{eq: identity})} which is satisfied exactly by the analytical expressions and verified numerically using finite-difference derivatives of \(\gamma_{\mathrm{num}}(\tau)\). Finally, it is crucial to emphasize that these validations are applied to the \emph{interconnected} kernels constructed from the bath spectral density. 

In \textbf{Figure \ref{fig: panel1}.}, coherent drive contributions were excluded from \(J_\mathrm{conn}(f)\) when forming \(\gamma(\tau)\), \(\nu(\tau)\), and \(\mu(\tau)\), and were instead treated separately through an explicit forcing term in the GLE. This separation ensures that the homogeneous dynamics—and hence the kernels—are identical in both the explicit-forcing and coherent-spectrum formulations, as demonstrated numerically. The panel shows both the classical and quantum corrections to the kernel contribute to the dynamics significantly at short timescales, but largely fall off. 

In panel \textbf{Figure \ref{fig: panel2}.} the coherent drive contributions were instead treated separately through the uncentered force sector at $F_0=0$. The equilibrium and reconstructed kernels admit the same distance from the stationary bath, underscoring their equivalence in alternate rotating frames.

\begin{figure}
    \centering

    \begin{overpic}[width=\linewidth]{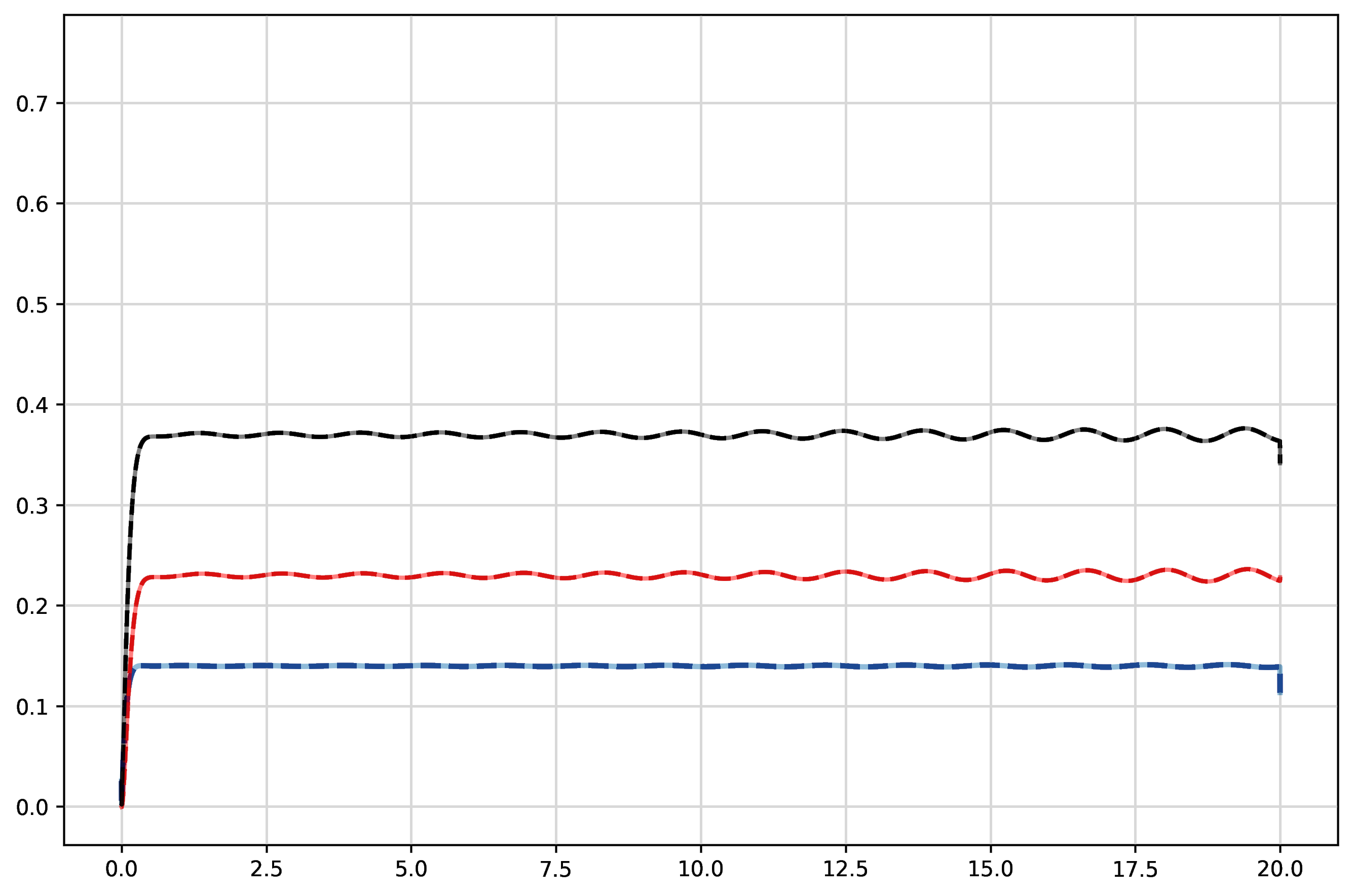}
        \put(7,58){\bfseries (a)}

        \put(53,55){
            \includegraphics[width=0.42\linewidth]{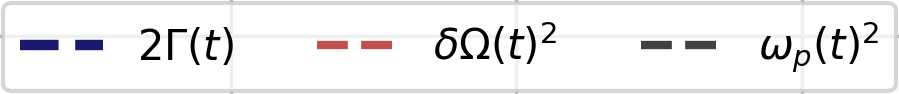}
        }
    \end{overpic}

    \vspace{0.5em}

    \begin{overpic}[width=\linewidth]{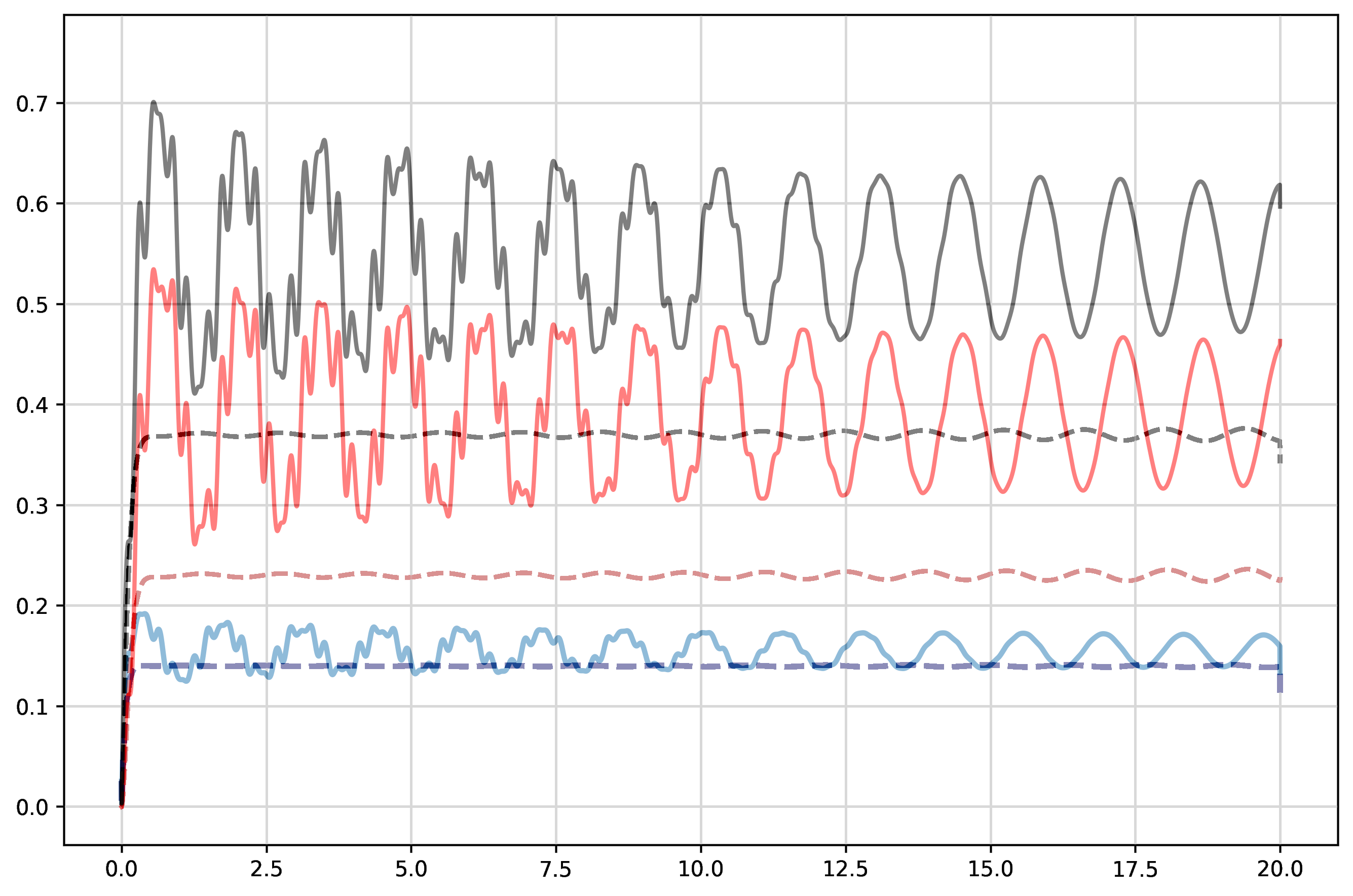}
        \put(7,58){\bfseries (b)}
    \end{overpic}

    \caption{\textit{Renormalization of the physically observable frequency under field bias.} Comparison between the equilibrium stationary renormalization \textbf{(a)} and the explicitly driven renormalization obtained in the laboratory frame \textbf{(b)}. The dashed curves denote the equilibrium contributions to the physically observable frequency, damping, and diffusion coefficients, while the solid oscillatory curves include the coherent field-bias contribution. In the driven representation, the oscillatory modulation appears as a persistent correction to the renormalized coefficients due to the explicit time dependence introduced by the coherent displacement. However, as demonstrated in the rotating-frame analysis of the preceding figures, these oscillations do not correspond to a modification of the underlying homogeneous dissipative dynamics. Instead, they arise from describing the same equilibrium kernel structure in a frame displaced by the coherent drive. The biased equilibrium and driven solutions therefore remain physically equivalent after transforming into the rotating frame of the coherent displacement. 
    }
    \label{fig:renormalization}
\end{figure}

\subsection{Discussion: \\
The physically observable frequency}

The code implements the generalized Langevin equation using a spectral representation of the bath, separating the total spectrum into
\begin{equation}
    J(\omega) = J_{\rm eq}(\omega) + J_{\rm inc}(\omega) + J_{\rm coh}(\omega).
\end{equation}
Only the connected part,
\begin{equation}
    J_{\rm conn}(\omega) = J_{\rm eq}(\omega) + J_{\rm inc}(\omega),
\end{equation}
is used to construct the memory kernel
\begin{equation}
    \gamma(t) = \frac{2}{\pi} \int_0^\infty d\omega\, \frac{J_{\rm conn}(\omega)}{\omega}\cos(\omega t),
\end{equation}
ensuring that the homogeneous dynamics—and therefore the physically observable frequency—are determined solely by fluctuating bath correlations.

The aforementioned simulations shown in \textbf{Figure \ref{fig: coherentforcing}.} confirm two key results. First, the GLE solution with explicit forcing is identical to that reconstructed from the Green’s function and the coherent spectral contribution, confirming that coherent drive produces only displacement. Second, the memory, dissipation, and noise kernels are unchanged between these descriptions, as can be noticed by their equivalent $L_2$ distances from the undriven equilibrium solution in \textbf{Figure \ref{fig: panel2}.}. 

To model the driven, nonstationary environment, the spectrum is promoted to a two-time (Wigner) form
\begin{equation}
    J_{\rm eff}(\omega;T) = J_{\rm eq}(\omega) + \mathcal{E}(T)\,J_{\rm inc}(\omega),
\end{equation}
which yields a center-time-dependent renormalization
\begin{equation}
    \Omega_{\rm ren}^{2}(T)
    =
    \Omega_{0}^{2}
    +
    \frac{2}{\pi}
    \int_{0}^{\infty} d\omega\, \frac{J_{\rm eff}(\omega;T)}{\omega}.
\end{equation}

The numerical benchmarks clarify an important conceptual point regarding the role of the coherent drive in the generalized Langevin equation and the associated field-biased HPZ master equation. 

Although the laboratory-frame kernels in the driven representation exhibit persistent oscillatory structure, these oscillations do not imply that the underlying dissipative dynamics have become intrinsically nonstationary. Rather, they arise because the coherent displacement introduces an explicitly time-dependent phase reference into the bath autocorrelation functions. In the equilibrium description, the bath kernels depend only on the relative time coordinate, so the corresponding spectral density is diagonal in frequency space and the reduced dynamics become stationary in the long-time limit. Consequently, the inhomogeneous Green's functions approach exponentially damped propagators with time-independent poles, and the dynamics admit the usual Markovian interpretation in which the bath memory decays on a finite correlation timescale.

When the coherent monochromatic displacement is introduced directly in the laboratory frame, the kernels acquire oscillatory prefactors proportional to the coherent phase evolution,
\begin{equation}
K_{\rm drv}(t,t')
\sim
e^{-i\omega_{\rm RF}(t-t')}
K_{\rm eq}(t-t'),
\end{equation}
which causes the corresponding renormalized frequency coefficients to oscillate indefinitely in the lab frame. However, these oscillations are inherited entirely from the externally imposed phase convention of the coherent drive and therefore do not correspond to an source of irreversible dynamics. 

Unsurprisingly, the coherent drive does not continually generate new dissipative structure in the connected bath correlations. The underlying memory kernel still relaxes to a fixed asymptotic form characterized by a finite bath correlation time,
$
\tau_B \sim \Omega_B^{-1},
$
and the renormalized coefficients approach bounded long-time values. The residual oscillations are therefore oscillations associated with the chosen rotating frame of the coherent field, not signatures of persistent nonstationary dissipation. Transforming into the rotating frame removes the explicit carrier-frequency dependence and makes this stationary behavior explicit:
$
K_{\rm rot}(t,t')
=
K_{\rm eq}(t-t').
$
The driven and undriven equilibrium kernels then become indistinguishable within numerical precision, confirming that both descriptions converge to the same stationary dissipative state. Consequently, both cases satisfy the conditions associated with Markovian behavior.


\section{Experimental Relevance}\label{sec: relevance}

Superconducting and circuit quantum electrodynamical (QED) hardware platforms have emerged as leading candidates for the realization of scalable quantum information processing \cite{devoret2013superconducting,blais2021circuitqed,kjaergaard2020superconducting}. 
These architectures provide a viable route toward universal quantum computation, as was demonstrated through high-fidelity gate operations and programmable processors \cite{barends2014logic,arute2019quantum}. 
In particular, bosonic encodings within superconducting circuits have enabled substantial progress toward hardware-efficient quantum error correction and fault tolerance \cite{ofek2016extending,hu2019quantum,cai2021bosonic,google2023errorcorrection}. 
Despite these major advances, decoherence—arising from intrinsic material losses and coupling to uncontrolled environmental degrees of freedom—remains the principal limitation to device performance \cite{clarke2008superconducting,martinis2005decoherence,burnett2019decoherence,wendin2017quantum,preskill2018nisq}. 

For example, continuously applied drives are essential for the manipulation and control of superconducting circuits \cite{Blais2004,Wallraff2004,DevoretSchoelkopf2013}. Dispersive readout \cite{Walter2017}, for instance, employs off-resonant drives to encode the system state in the phase of a probe electromagnetic field \cite{Blais2004,Clerk2010}. More generally, continuous drives offer a critical tool for pushing quantum hardware to its operational limits, enabling engineered dissipation via controlled decay channels activated through nonlinear mixing processes \cite{Shruti, Zaki, Jeff}, each instance leading to the experimental observation of thermal effects not accounted for by the Markovian model, and well-documented to be caused by the continuously applied off-resonant fields across the literature. Moreover, the theory of quantum noise and open quantum systems has historically been dominated by models based on harmonic (quadratic) potentials, due to their exact solvability and Gaussian structure \cite{breuer2002theory,gardiner2004quantum,weiss2012quantum,caldeira1983path,hu1992quantum}. 
However, it is well established that such strictly linear (Gaussian) systems are insufficient for universal quantum computation, as they can be efficiently simulated classically and lack the necessary resources for selectively exciting transitions \cite{lloyd1999quantum,braunstein2005quantum,weedbrook2012gaussian}. 
This fundamental limitation underscores the necessity of incorporating nonlinear elements—such as anharmonic potentials or effective interactions—in order to achieve computational universality \cite{nielsen2000quantum,krantz2019quantum}. 

Thus, a complete understanding of quantum noise in realistic quantum devices requires going beyond the harmonic potential. To this end, we conclude this work by presenting two potential extensions to the present framework directly applicable to the language of superconducting quantum circuits.

\subsection{Representing the circuit QED Hamiltonian}

The subsystem-environment construction where the classical drive $E(t)$ acts on both the subsystem and reservoir degrees of freedom can be generalized to an ensemble of $N_i$-many interacting subsystems indexed by $i$. Each subsystem coordinate $x_i$ is coupled to its own continuum of harmonic oscillators $x_{ij}$, The total Hamiltonian is of the form
\begin{align}
H = &\sum_i^{N_i} H_S^{(i)}(x_i,p_i) 
- \sum_i^{N_i} x_i\,E(t)
  \\&+ \sum_{i}^{N_i}\sum_{j} \Big(H_{B}^{(ij)}(x_{ij},p_{ij}) - x_{ij}E(t) + c_{ij}\, x_{ij}\, x_i \Big). \nonumber
\end{align}
with frequencies $\omega_{ij}$ and bilinear coupling constants $c_{ij}$ characterizing the interaction between subsystem coordinate $x_i$ and the $j$th oscillator of its reservoir. Each bath oscillator $j$ associated with subsystem $i$ is taken as CL-type reservoir \cite{CaldeiraLeggett},
\begin{equation}
H_{B}^{(ij)}(x_{ij},p_{ij}) =
\frac{p_{ij}^2}{2m_{ij}} + \frac{1}{2} m_{ij}\omega_{ij}^2 x_{ij}^2,
\end{equation}
The Hamiltonian for the subsystem ensemble is 
another set of harmonic oscillators,
\begin{equation}
H_S^{(i)}(x_i,p_i) = \frac{p_i^2}{2m_i} + \frac{1}{2} m_i \omega_i^2 x_i^2 + V(x_i, x_{i'}),
\end{equation}
where the potential $V(x_i,x_{i'})$ contains any additional sub-harmonic perturbations or inter-coupling between distinct subsystem position coordinates, $x_i,\,x_{i'}$. In the particular context of superconducting circuits, further specifications about the form of $V(x_i,x_{i'})$ can be made.  

 
The superconducting Josephson element contributes a potential $V(\varphi)=E_J \cos(\varphi) -\varphi^2/2$, where the zero-point mode fluctuations operator $\varphi$ capacitively couples the sum of linearized resonator modes, leading to a inter-coupling potential of the form \cite{Wallraff2004}
\begin{equation}
\label{eq: potential}
    V(x_i, x_{i'}) = E_J \cos
\Big(\sum_i^{N_i}  
    \dfrac{x_i}{\tilde{\varphi}_i}\Big) -\dfrac{1}{2}
    \Big(\sum_i^{N_i}  
    \dfrac{x_i}{\tilde{\varphi}_i}\Big)
\end{equation}
or, equivalently, under a Taylor series expansion
\begin{equation}
\label{eq:Taylorpotential}
    V(x_i, x_{i'}) \propto E_J \sum_{n=2}^{\infty}
\frac{(-1)^n}{(2n)!}\Big(\sum_i^{N_i}  
    \dfrac{x_i}{\tilde{\varphi}_i}\Big)^{2n}.
\end{equation}
It should be noted that while $N_i\geq1$ includes $N_i=1$, the junction contributes a linear mode that dresses the rest of the system frequencies s.~t. $N_i\rightarrow 1$, and for that matter, most systems require $N_i=2$ or greater to be able to conduct indirect measurements on the information-bearing (deemed the storage) mode \cite{DevoretSchoelkopf2013}. 



Eliminating the reservoir degrees of freedom at the operator level yields a GLE for each subsystem coordinate $x_i(t)$. The resulting intercoupled Heisenberg EoM for the $i$-th assumes the form
\begin{align}
M_i \ddot{x}_i(t)
+&
M_i \bar{\Omega}_i^2 x_i(t)
+
\frac{\partial V}{\partial x_i}
\\&+
\int_{0}^{t} ds\, \gamma_i(t-s)\,\dot{x}_i(s)
=
\tilde{F}_i(t),
\nonumber 
\end{align}
where $\gamma_i(t)$ is the $i$-th bath {memory} kernel, and $\bar{\Omega}_i$ is not the bare frequency $\bar{\omega}_i$, but rather the bath-renormalized subsystem frequency. The operator-valued fluctuating force, $\tilde{F}_i(t)$,
contains both the intrinsic reservoir fluctuations $F_i(t)$ and the additional fluctuations $F_{i}^{E}(t)$ induced by the external driving field $E(t)$ acting on the reservoir modes. 

The perturbation $V(x_i, x_i')$ leaves the kernels unchanged, but introduces modifications to the diffusion coefficients of the master equation. While, in general, these corrections to the coefficients should not be considered (analytically) tractable, the boundedness of the Josephson potential in \textbf{Eq.~(\ref{eq: potential})} allows us to defer the problem to numerics at the Green's function stage by means of a perturbative expansion.

\subsection{Representing coherent driven interactions}

One immediate extension to consider in the present framework is the time-varying bichromatic field. Physically, this corresponds to the application of a coherent RF drive in circuit QED, where the (continuously) applied drive tone is routinely used to control, stabilize, and/or probe superconducting quantum devices in the rotating frame of a reference frequency we can regard as the "beat" tone. Starting from a two-tone instantaneous field \cite{Zaki, blais2021circuitqed},
\begin{align}
\label{field}
E(t)
=& 2\Re\{\varepsilon_p e^{-i\omega_p t}
+
\varepsilon_d e^{-i\omega_d t}\} 
\end{align}
where $\varepsilon_p$ and $\varepsilon_d$ are arbitrary complex amplitudes, and no symmetry between the two tones is assumed, and we have rewritten the fields in \textbf{Eq. (\ref{field})} by expanding the real part using the identity
$
2\mathrm{Re}{\{z\}} = (z + z^*),
$
thereby expressing the signal as a sum of the positive- and negative-frequency components.  It is convenient to introduce the central frequency $\omega_c$ and half-difference (carrier detuning) $\Delta_c$ frequency, 
\begin{equation}
\omega_c = \frac{\omega_p+\omega_d}{2},
\qquad
\Delta_c = \frac{\omega_p-\omega_d}{2},
\end{equation}
so that with
$
\omega_p = \omega_c + \Delta_c,
$ and $
\omega_d = \omega_c - \Delta_c,
$
the field may then be rewritten as
\begin{align}
E(t)
=\; &
e^{-i\omega_c t}
(
\varepsilon_p e^{-i\Delta_c t}
+
\varepsilon_d e^{+i\Delta_c t}
)
\\&\quad+
e^{i\omega_c t}
(
\varepsilon_p^{*} e^{i\Delta_c t}
+
\varepsilon_d^{*} e^{-i\Delta_c t}
).
\end{align}

Defining carrier envelope,
$
\mathcal{E}_{c}(t)
=
\varepsilon_p e^{-i\Delta_c t}
+
\varepsilon_d e^{i\Delta_c t}
$,
the drive can be rewritten as
\begin{equation}
E(t)
=
\mathcal{E}_{c}(t)e^{-i\omega_c t}
+
\mathcal{E}_{c}^{*}(t)e^{i\omega_c t}.
\end{equation}
In this representation, the \enquote{beat} tone is encoded in the slowly-varying envelope, $\mathcal{E}_{c}(t)$, while the rapid oscillation is carried by the central frequency $\omega_c$.
Later on, it can also be helpful to define the average carrier complex amplitude and amplitude imbalance,
\begin{equation}
\bar{\mathcal{E}}_c
=
\frac{\varepsilon_p+\varepsilon_d}{2},
\qquad
\delta \mathcal{E}_c
=
\frac{\varepsilon_p-\varepsilon_d}{2},
\end{equation}
in terms of which
\begin{equation}
\dfrac{\mathcal{E}_{c}(t)}{2}
=
\mathcal{E}_c\cos(\Delta_c t)
-
i\,\delta\bar{\mathcal{E}}_c \sin(\Delta_c t).
\end{equation}
Thus, the two-tone field is exactly equivalent to a single carrier at frequency $\omega_c$ with a time-dependent envelope. Accordingly, the full two-time correlator, $E(t)E(t')$, may be written in a compact carrier form, resulting in the expression
\begin{align}
\label{eq: noRWA}
E(t)E(t')
&=\;
2\,\mathrm{Re}
(
\mathcal{E}_{c}(t)\mathcal{E}_{c}(t')e^{-i\omega_c(t+t')}
)
\\&\quad+
2\,\mathrm{Re}
(
\mathcal{E}_{c}(t)\mathcal{E}_{c}^{*}(t')e^{-i\omega_c(t-t')}
).\nonumber
\end{align}
The first term depends on the sum time $t+t'$ and represents the (rapidly oscillating) counter-rotating contribution (in the rotating frame of $\omega_c$), whereas the second depends on both the carrier difference time $t-t'$, and $\mathcal{E}_{c}(t)\mathcal{E}_{c}^{*}(t')$ contains the beat tone modulation. 

If one then further performs the RWA with respect to the carrier frequency $\omega_c$, the counter-terms (proportional to ${\pm i\omega_c(t+t')}$) are dropped, leaving the (reduced) two-time correlator
\begin{equation}
\label{eq: RWA}
E(t)E(t')
\approx
2\,\mathrm{Re}
\!\left[
\mathcal{E}_{c}(t)\mathcal{E}_{c}^{*}(t')e^{-i\omega_c(t-t')}
\right].
\end{equation}

Here, we can appreciate two things about the carrier envelope notation we have introduced. First, taking a ($L_2$-norm) distance measure between \textbf{Eq.~(\ref{eq: RWA})} and \textbf{Eq.~(\ref{eq: noRWA})} recovers the error incurred in the two-time field AoC from the application of the RWA. Second, the carrier central frequency and detuning frequency (equivalently, the carrier envelope) can be re-defined compactly for an arbitrary number of applied tones.

Consider a $D_j$-tone set of $D_k$-many frequency combs, applied to $D_i$-many central frequencies $\in \{\Omega_i\}$, %
\begin{equation}
\label{eq: FWM}
E(t) = 
\sum_{j=1}^{D_j}
\sum_{k=1}^{D_k}
\mathrm{Re}
\{\,
\sum_{-D_i}^{D_i}
\tilde{\mathcal{E}}_{j i k}\,
e^{-i \Omega_{k} t}
\,\},
\end{equation}
following the convention for  $\tilde{\Omega}_i:=\left(\Omega_i + \Delta_i + i|\,\Delta_i^{S}\,|\right)
$, wherein $\Delta_i$ denotes the detunings from the $\Omega_i$-th atomic transition frequency with (frequency-dependent) comb spacings $\Delta_i^S$, where equal spacings $\{\Delta_i^S\}\rightarrow |\,\Delta_i^{S}\,|$ are routinely used to selectively excite multi-level transitions in quantum optics \cite{AQST, Shruti, Jeff}. 

For each drive tone, it is useful to separate the applied frequency into a carrier component and a detuning from that carrier.  Thus, for the poly-chromatic tone denoted by the triple index $(j,k,\ell)$, we define
\begin{align}
\omega_{c,jk\ell}
&=
\Omega_{j}
+
\Delta_{k},
\\
\Delta_{c,jk\ell}
&=
\Omega_{j}
+
\Delta_{k}
+
\left|\Delta_{\ell}^{S}\right|
-
\omega_{c,jk\ell}
=
\left|\Delta_{\ell}^{S}\right|.
\end{align}
Equivalently, the field in \textbf{Eq. (\ref{eq: FWM})} becomes
\begin{align}
\label{eq: FWMfield2}
E(t)
&=
\sum_{j=1}^{D_{\Omega}}
\sum_{k=1}^{D_{\Delta}}
\mathrm{Re}
\big\{
e^{-i\omega_{jk\ell}t}
\sum_{\ell=1}^{D_{S}}
\tilde{E}_{jk\ell}
e^{-i\Delta_{jk\ell}t}
\big\},
\end{align}
wherein $\omega_{jk\ell}$ is the central carrier frequency associated with the $(j,k)$ transition-and-detuning pair, while $\Delta_{jk\ell}$ represents the residual carrier detuning induced by the $\ell$-th spectral-spacing component.

To place \textbf{Eq.~(\ref{eq: FWMfield2})} in a compacted carrier form analogous to the bi-chromatic decomposition, we factor out the common carrier frequency and collect all of the residual frequency components into a general carrier envelope. Selecting the reference central frequency $\omega_c$, the field may be rewritten as
\begin{align}
E(t)
&=
\mathrm{Re}
\left\{
e^{-i\omega_c t}
\mathcal{E}_c(t)
\right\},
\end{align}
where the carrier envelope $\mathcal{E}_c(t)$ is defined by
\begin{align}
\mathcal{E}_c(t)
&=
\sum_{j=1}^{D_{\Omega}}
\sum_{k=1}^{D_{\Delta}}
\sum_{\ell=1}^{D_{S}}
\tilde{E}_{jk\ell}\,
e^{-i\delta_{jk\ell} t},
\end{align}
with detunings
$
\delta_{jk\ell}
=
\Omega_j + \Delta_k + \Delta_\ell^{S} - \omega_c
$. In this representation, the rapidly oscillating component is carried by the carrier central frequency $\omega_c$, while all relative frequency offsets contribute to the slow modulation encoded in $\mathcal{E}_c(t)$. This decomposition generalizes the two-tone carrier–envelope form to an arbitrary multi-frequency comb, with the envelope capturing the full set of amplitudes.

Lastly, it is worth stating when this generalization to multichromatic is applicable for the quantum mechanical treatment of the drive. In general, we have assumed that the quantum mechanical field bias originates from a single quanta, which negates cross-correlations in the fluctuations between drives applied from different sources. Where this approximation holds remains an active area of research directly pertinent to multiplexing in scaled processors, where drive crosstalk becomes a significant source of noise.


\section{Conclusions}

In this work, we developed a modified quantum master equation for driven open quantum systems by extending the CL formalism to include explicit field bias and cross field-bath couplings. Starting from the exact elimination of a Gaussian reservoir, we derived a field-biased GLE and the corresponding modified HPZ master equation with explicitly time-dependent noise, dissipation, and renormalization kernels which ultimately become stationary. By comparing equilibrium, classically driven, and quantum-driven treatments on equal footing, we demonstrated that coherent displacements of the sub-system can be incorporated into the open quantum system in two formally distinct but physically equivalent ways: either as an explicit forcing term in the GLE, or through a renormalization of the system Hamiltonian through a nonstationary reservoir. At the operator level of the sub-system dynamics, these two descriptions are completely equivalent. 

In both cases, the physically observable resonant frequency remains determined by the pole structure of the homogeneous Green's function associated with the dissipation kernel. When the external field is treated classically, the dissipation kernel is unchanged, and the coherent displacement enters solely as an additive forcing term. In a fully quantum treatment, the same displacement can be absorbed into a re-definition of the reservoir operators at the (explicitly two-time) continuum limit, yielding an equivalent description in which the coherent response is encoded in a shift of the first and second moments of the canonical position and momentum operators. 

As expected, the major distinction between the classical and quantum mechanical treatment of the drive arises in the fluctuation sector. In the classical treatment, the field contributes an additional term to the noise kernel proportional to the product of the field at two times, leading to explicitly nonstationary contributions which cannot modify the underlying bath spectral density, since they do not commute. In contrast, when the field is treated quantum mechanically, the noise correlations are determined by operator expectation values that incorporate both coherent and incoherent components of the field. For a purely coherent state, the connected part of the noise kernel remains identical to that of the equilibrium bath, and the field-bias induced contribution reduces to the same factorized form obtained in the classical field limit. In this sense, the classical and quantum treatments are equivalent for coherent driving at the level of fluctuations, once the separation between mean displacement and connected correlations is properly accounted for.

Regardless of whether the classical treatment of the drive is feasible, only the quantum mechanical treatment has the potential to enter the coherent dynamics as a mass renormalization of the physically observable frequency. Deliberately doing so presents two potential advantages in the emulation of noise in quantum hardwares which have not yet been demonstrated: \textbf{(1)} improvements in error scaling rates for perturbative corrections (as opposed to the classical drive treatment) to coherent displacement for nonlinear systems \cite{breuer2002theory} and \textbf{(2)} the unique opportunity to study (and therefore, optimize around) how pulse shapes and duration bias the thermal noise in a quantum processor. 

In addition to presenting and demonstrating the accuracy and convergence to the classical limit of our field-biased quantum master equation, we presented the avenues for accomplishing both, laying the foundation for future work. As such, our field-biased HPZ quantum master equation offers a unified microscopic foundation that not only reconciles the classical and quantum limits of continuously driven open quantum systems, but also furnishes a scalable tool for analyzing and engineering non-Markovian effects in complex quantum environments directly accessible in the language of superconducting and/or driven-dissipative quantum circuits.



\section*{Acknowledgements} \label{sec:acknowledgements}

This material is based upon work supported by the U.S. Department of Energy, Office of Science, Office of Workforce Development for Teachers and Scientists, Office of Science Graduate Student Research (SCGSR) program. The SCGSR program is administered by the Oak Ridge Institute for Science and Education (ORISE) for the DOE. M. G. B thanks her experimentalist and device theory colleagues for their feedback and insight on applicability to superconducting hardwares: Shruti Shirol, Lev-Arcady Sellem, Nicolas Dirnegger and Cody Fan in particular for helpful discussions. 

\nocite{*}
\bibliographystyle{unsrt}   
\bibliography{citations}

@book{breuer2002theory,
  title={The Theory of Open Quantum Systems},
  author={Breuer, H.-P. and Petruccione, F.},
  publisher={Oxford University Press},
  year={2002}
}

@article{moyal1949,
  author = {J. E. Moyal},
  title = {Quantum Mechanics as a Statistical Theory},
  journal = {Mathematical Proceedings of the Cambridge Philosophical Society},
  volume = {45},
  number = {1},
  pages = {99--124},
  year = {1949},
  doi = {10.1017/S0305004100000487}
}

@book{poisson1809,
  author = {Sim{\'e}on Denis Poisson},
  title = {Trait{\'e} de M{\'e}canique},
  publisher = {Bachelier},
  address = {Paris},
  year = {1809}
}

@book{liouville1838,
  author = {Joseph Liouville},
  title = {Journal de Math{\'e}matiques Pures et Appliqu{\'e}es},
  publisher = {Bachelier},
  address = {Paris},
  volume = {3},
  pages = {342--349},
  year = {1838},
  note = {Original paper introducing Liouville's theorem}
}

@phdthesis{bishop2010cqed,
  author       = {Bishop, Lev S.},
  title        = {Circuit Quantum Electrodynamics},
  school       = {Yale University},
  year         = {2010},
  month        = {May},
  note         = {PhD thesis},
  eprint       = {1007.3520},
  archivePrefix= {arXiv},
  primaryClass = {quant-ph},
  url          = {https://arxiv.org/abs/1007.3520}
}

@book{pozar2012microwave,
  author    = {Pozar, David M.},
  title     = {Microwave Engineering},
  edition   = {4},
  publisher = {John Wiley \& Sons},
  address   = {Hoboken, NJ},
  year      = {2012},
  isbn      = {9780470631553}
}

@article{Pedestrians,
  author  = {W. B. Case},
  title   = {Wigner functions and Weyl transforms for pedestrians},
  journal = {American Journal of Physics},
  volume  = {76},
  number  = {10},
  pages   = {937--946},
  year    = {2008},
  doi     = {10.1119/1.2957889}
}

@article{caldeira1983path,
  title={Path Integral Approach to Quantum Brownian Motion},
  author={Caldeira, A. O. and Leggett, A. J.},
  journal={Physica A},
  volume={121},
  pages={587--616},
  year={1983}
}

@article{hu1992quantum,
  title={Quantum Brownian Motion in a General Environment: Exact Master Equation},
  author={Hu, B. L. and Paz, J. P. and Zhang, Y.},
  journal={Physical Review D},
  volume={45},
  pages={2843--2861},
  year={1992}
}

@book{gardiner2004quantum,
  title={Quantum Noise},
  author={Gardiner, C. W. and Zoller, P.},
  publisher={Springer},
  year={2004}
}

@article{wallraff2004,
  author = {Wallraff, A. and Schuster, D. I. and Blais, A. and Frunzio, L. and Huang, R.-S. and Majer, J. and Kumar, S. and Girvin, S. M. and Schoelkopf, R. J.},
  title = {Strong coupling of a single photon to a superconducting qubit using circuit quantum electrodynamics},
  journal = {Nature},
  volume = {431},
  pages = {162--167},
  year = {2004}
}

@article{strongcoupling_blais2004,
  author = {Blais, A. and Huang, R.-S. and Wallraff, A. and Girvin, S. M. and Schoelkopf, R. J.},
  title = {Cavity quantum electrodynamics for superconducting electrical circuits: An architecture for quantum computation},
  journal = {Phys. Rev. A},
  volume = {69},
  pages = {062320},
  year = {2004}
}

@article{koch2007transmon,
  author = {Koch, J. and Yu, T. M. and Gambetta, J. and Houck, A. A. and Schuster, D. I. and Majer, J. and Blais, A. and Devoret, M. H. and Girvin, S. M. and Schoelkopf, R. J.},
  title = {Charge-insensitive qubit design derived from the Cooper pair box},
  journal = {Phys. Rev. A},
  volume = {76},
  pages = {042319},
  year = {2007}
}

@article{paik2011observation,
  author = {Paik, H. and Schuster, D. I. and Bishop, L. S. and Kirchmair, G. and Catelani, G. and Sears, A. P. and Johnson, B. R. and Reagor, M. and Frunzio, L. and Glazman, L. I. and Girvin, S. M. and Devoret, M. H. and Schoelkopf, R. J.},
  title = {Observation of high coherence in Josephson junction qubits measured in a three-dimensional circuit QED architecture},
  journal = {Phys. Rev. Lett.},
  volume = {107},
  pages = {240501},
  year = {2011}
}

@article{reagor2016quantum,
  author = {Reagor, M. and Pfaff, W. and Axline, C. and Heeres, R. and Ofek, N. and Sliwa, K. M. and Holland, E. and Wang, C. and Blumoff, J. and Chou, K. and others},
  title = {Quantum memory with millisecond coherence in circuit QED},
  journal = {Phys. Rev. B},
  volume = {94},
  pages = {014506},
  year = {2016}
}

@article{gertler2021protecting,
  author = {Gertler, J. M. and Zou, C.-L. and Schuster, D. I. and Girvin, S. M. and Jiang, L.},
  title = {Protecting a bosonic qubit with autonomous quantum error correction},
  journal = {Nature},
  volume = {590},
  pages = {243--248},
  year = {2021}
}

@book{gibbs1902,
  author    = {J. W. Gibbs},
  title     = {Elementary Principles in Statistical Mechanics: Developed with Especial Reference to the Rational Foundation of Thermodynamics},
  publisher = {Charles Scribner's Sons},
  address   = {New York},
  year      = {1902},
  note      = {See Chapter IV: "On the Distribution in Phase Called Canonical"}
}

@book{weiss2012quantum,
  title={Quantum Dissipative Systems},
  author={Weiss, U.},
  publisher={World Scientific},
  year={2012}
}

@article{lloyd1999quantum,
  title={Quantum Computation over Continuous Variables},
  author={Lloyd, S. and Braunstein, S. L.},
  journal={Physical Review Letters},
  volume={82},
  pages={1784--1787},
  year={1999}
}

@article{braunstein2005quantum,
  title={Quantum Information with Continuous Variables},
  author={Braunstein, S. L. and van Loock, P.},
  journal={Reviews of Modern Physics},
  volume={77},
  pages={513--577},
  year={2005}
}

@article{weedbrook2012gaussian,
  title={Gaussian Quantum Information},
  author={Weedbrook, C. et al.},
  journal={Reviews of Modern Physics},
  volume={84},
  pages={621--669},
  year={2012}
}

@article{bartlett2002classical,
  title={Classical and Quantum Communication without a Shared Reference Frame},
  author={Bartlett, S. D. and Sanders, B. C.},
  journal={Physical Review Letters},
  volume={89},
  pages={207903},
  year={2002}
}

@article{nielsen2000quantum,
  title={Quantum Computation and Quantum Information},
  author={Nielsen, M. A. and Chuang, I. L.},
  journal={Cambridge University Press},
  year={2000}
}

@article{krantz2019quantum,
  title={A Quantum Engineer's Guide to Superconducting Qubits},
  author={Krantz, P. et al.},
  journal={Applied Physics Reviews},
  volume={6},
  pages={021318},
  year={2019}
}

@article{devoret2013superconducting,
  title={Superconducting Circuits for Quantum Information: An Outlook},
  author={Devoret, M. H. and Schoelkopf, R. J.},
  journal={Science},
  volume={339},
  pages={1169--1174},
  year={2013}
}

@article{blais2021circuitqed,
  title={Circuit Quantum Electrodynamics},
  author={Blais, A. and Grimsmo, A. L. and Girvin, S. M. and Wallraff, A.},
  journal={Reviews of Modern Physics},
  volume={93},
  pages={025005},
  year={2021}
}

@article{kjaergaard2020superconducting,
  title={Superconducting Qubits: Current State of Play},
  author={Kjaergaard, M. and Schwartz, M. E. and Braum{\"u}ller, J. and Krantz, P. and Wang, J. I.-J. and Gustavsson, S. and Oliver, W. D.},
  journal={Annual Review of Condensed Matter Physics},
  volume={11},
  pages={369--395},
  year={2020}
}

@article{arute2019quantum,
  title={Quantum Supremacy Using a Programmable Superconducting Processor},
  author={Arute, F. et al.},
  journal={Nature},
  volume={574},
  pages={505--510},
  year={2019}
}

@article{barends2014logic,
  title={Logic Gates at the Surface Code Threshold},
  author={Barends, R. et al.},
  journal={Nature},
  volume={508},
  pages={500--503},
  year={2014}
}

@article{preskill2018nisq,
  title={Quantum Computing in the NISQ Era and Beyond},
  author={Preskill, J.},
  journal={Quantum},
  volume={2},
  pages={79},
  year={2018}
}

@article{ofek2016extending,
  title={Extending the Lifetime of a Quantum Bit with Error Correction in Superconducting Circuits},
  author={Ofek, N. et al.},
  journal={Nature},
  volume={536},
  pages={441--445},
  year={2016}
}

@article{hu2019quantum,
  title={Quantum Error Correction and Universal Gate Set Operation on a Binomial Bosonic Logical Qubit},
  author={Hu, L. et al.},
  journal={Nature Physics},
  volume={15},
  pages={503--508},
  year={2019}
}

@article{cai2021bosonic,
  title={Bosonic Quantum Error Correction Codes in Superconducting Circuits},
  author={Cai, W. and Ma, Y. and Wang, W. and Zou, C.-L. and Sun, L.},
  journal={Fundamental Research},
  volume={1},
  pages={50--67},
  year={2021}
}

@article{google2023errorcorrection,
  title={Suppressing Quantum Errors by Scaling a Surface Code Logical Qubit},
  author={Google Quantum AI},
  journal={Nature},
  volume={614},
  pages={676--681},
  year={2023}
}

@article{clarke2008superconducting,
  title={Superconducting Quantum Bits},
  author={Clarke, J. and Wilhelm, F. K.},
  journal={Nature},
  volume={453},
  pages={1031--1042},
  year={2008}
}

@article{martinis2005decoherence,
  title={Decoherence in Josephson Qubits from Dielectric Loss},
  author={Martinis, J. M. et al.},
  journal={Physical Review Letters},
  volume={95},
  pages={210503},
  year={2005}
}

@article{burnett2019decoherence,
  title={Decoherence Benchmarking of Superconducting Qubits},
  author={Burnett, J. et al.},
  journal={npj Quantum Information},
  volume={5},
  pages={54},
  year={2019}
}

@article{wendin2017quantum,
  title={Quantum Information Processing with Superconducting Circuits},
  author={Wendin, G.},
  journal={Reports on Progress in Physics},
  volume={80},
  pages={106001},
  year={2017}
}

@misc{Fibre,
  author       = {Billington, R.},
  title        = {A Report on Four-Wave Mixing in Optical Fibre and its Metrological Applications},
  institution  = {National Physical Laboratory},
  number       = {NPL Report COEM 24},
  year         = {1999},
  note         = {NPL Report},
  url          = {https://eprintspublications.npl.co.uk/1036/1/coem24.pdf}
}

@book{matsubara1955,
  author = {T. Matsubara},
  title = {A New Approach to Quantum-Statistical Mechanics},
  journal = {Progress of Theoretical Physics},
  volume = {14},
  number = {4},
  pages = {351--378},
  year = {1955},
  doi = {10.1143/PTP.14.351},
  publisher = {cambridge university press}
}

@incollection{FWMintro,
  author    = {Thyagarajan, K. and Ghatak, A. K.},
  title     = {Fiber and Guided Wave Optics | Nonlinear Optics},
  booktitle = {Encyclopedia of Modern Optics},
  pages     = {472--486},
  year      = {2005},
  publisher = {Elsevier},
  url       = {https://www.sciencedirect.com/topics/physics-and-astronomy/four-wave-mixing}
}

@article{Lindblad1976,
  author = {Lindblad, G{\"o}ran},
  title = {On the Generators of Quantum Dynamical Semigroups},
  journal = {Communications in Mathematical Physics},
  volume = {48},
  pages = {119--130},
  year = {1976},
  doi = {10.1007/BF01608499}
}

@article{Gorini1976,
  author = {Gorini, Vittorio and Kossakowski, Andrzej and Sudarshan, E. C. G.},
  title = {Completely Positive Dynamical Semigroups of N-Level Systems},
  journal = {Journal of Mathematical Physics},
  volume = {17},
  pages = {821--825},
  year = {1976},
  doi = {10.1063/1.522979}
}

@article{Brasil2012,
  author = {Brasil, Carlos Alexandre and Fanchini, Felipe Fernandes and Napolitano, Reginaldo de Jesus},
  title = {A simple derivation of the Lindblad equation},
  journal = {arXiv preprint arXiv:1110.2122},
  year = {2012},
  archivePrefix = {arXiv},
  eprint = {1110.2122},
  primaryClass = {quant-ph},
  url = {https://arxiv.org/abs/1110.2122}
}

@article{Crystal,
  author  = {Powell, R. C. and Payne, S. A.},
  title   = {Dispersion effects in four-wave mixing measurements of ions in solids},
  journal = {Optics Letters},
  volume  = {15},
  pages   = {1233--1235},
  year    = {1990},
  url     = {https://opg.optica.org/ol/abstract.cfm?URI=ol-15-21-1233}
}

@inproceedings{Gas,
  author    = {Pooser, R. C. and Boyer, V. and Marino, A. M. and Lett, P. D.},
  title     = {Squeezed Light and Entangled Images from Four-Wave-Mixing in Hot Rubidium Vapor},
  booktitle = {Quantum Communications and Quantum Imaging VI},
  year      = {2008},
  address   = {San Diego, CA, USA},
  url       = {https://arxiv.org/pdf/0811.1243}
}

@phdthesis{TWPAthesis,
  author  = {Schackert, F. D. O.},
  title   = {A Practical Quantum-Limited Parametric Amplifier Based on the Josephson Ring Modulator},
  school  = {Yale University},
  year    = {2013},
  month   = {December},
  url     = {https://bpb-us-w2.wpmucdn.com/campuspress.yale.edu/dist/2/3627/files/2021/07/Thesis_Flavius_A-Practical-Quantum-Limited-Parametric-Amplifier-2.pdf}
}

@article{TWPApaper,
  author  = {Bergeal, N. and Schackert, F. and Metcalfe, M. and Vijay, R. and Manucharyan, V. E. and Frunzio, L. and Prober, D. E. and Schoelkopf, R. J. and Girvin, S. M. and Devoret, M. H.},
  title   = {Phase-preserving amplification near the quantum limit with a Josephson ring modulator},
  journal = {Nature},
  volume  = {465},
  pages   = {64--68},
  year    = {2010},
  url     = {https://proberlab.yale.edu/sites/default/files/nature09035.pdf}
}

@article{Zaki,
  author       = {Z. Leghtas and S. Touzard and I. M. Pop and A. Kou and B. Vlastakis and A. Petrenko and K. M. Sliwa and A. Narla and S. Shankar and M. J. Hatridge and M. Reagor and L. Frunzio and R. J. Schoelkopf and M. Mirrahimi and M. H. Devoret},
  title        = {Confining the state of light to a quantum manifold by engineered two-photon loss},
  journal      = {Science},
  volume       = {347},
  number       = {6224},
  pages        = {853--857},
  year         = {2015},
  doi          = {10.1126/science.aaa2085},
  eprint       = {1412.4633},
  archivePrefix= {arXiv},
  primaryClass = {quant-ph}
}

@article{TWPAseminal,
  author  = {Yurke, B. and Kaminsky, P. G. and Miller, R. E. and Whittaker, E. A. and Smith, A. D. and Silver, A. H. and Simon, R. W.},
  title   = {Observation of parametric amplification and deamplification in a Josephson parametric amplifier},
  journal = {Physical Review A},
  volume  = {39},
  pages   = {2519},
  year    = {1989},
  url     = {https://journals.aps.org/pra/pdf/10.1103/PhysRevA.39.2519}
}

@article{Alfonso,
  author  = {Kim, S. and Marino, A. M.},
  title   = {Generation of {${}^{87}$}Rb resonant bright two-mode squeezed light with four-wave mixing},
  journal = {Optics Express},
  volume  = {26},
  pages   = {33366--33375},
  year    = {2018},
  doi     = {10.1364/OE.26.033366},
  url     = {https://doi.org/10.1364/OE.26.033366}
}

@article{Jeff,
  author  = {Gertler, J. M. and Baker, B. and Li, J. and Shirol, S. and Koch, J. and Wang, C.},
  title   = {Protecting a bosonic qubit with autonomous quantum error correction},
  journal = {Nature},
  volume  = {590},
  pages   = {243--248},
  year    = {2021},
  url     = {https://arxiv.org/pdf/2004.09322}
}

@article{Shruti,
  author  = {Shirol, S. and van Gelderen, S. and Xi, H. and Wang, C.},
  title   = {Passive quantum error correction of photon loss at breakeven},
  journal = {arXiv},
  year    = {2025},
  eprint  = {2510.19794},
  archivePrefix = {arXiv},
  url     = {https://arxiv.org/pdf/2510.19794}
}

@article{Axions,
  author  = {Bartram, C. and et al.},
  title   = {Dark matter axion search using a Josephson traveling wave parametric amplifier},
  journal = {Review of Scientific Instruments},
  volume  = {94},
  pages   = {044703},
  year    = {2023},
  url     = {https://arxiv.org/abs/2110.10262}
}

@article{AQST,
  author  = {Wang, C. and Gertler, J. M.},
  title   = {Autonomous quantum state transfer by dissipation engineering},
  journal = {Physical Review Research},
  volume  = {1},
  pages   = {033198},
  year    = {2019},
  url     = {https://journals.aps.org/prresearch/pdf/10.1103/PhysRevResearch.1.033198}
}

@book{BreuerPetruccione,
  author    = {Breuer, H.-P. and Petruccione, F.},
  title     = {The Theory of Open Quantum Systems},
  publisher = {Oxford University Press},
  address   = {Oxford},
  year      = {2002}
}

@book{GardinerZoller,
  author    = {Gardiner, C. W. and Zoller, P.},
  title     = {Quantum Noise},
  publisher = {Springer},
  address   = {Berlin},
  year      = {2004}
}

@article{CaldeiraLeggett,
  author  = {Caldeira, A. O. and Leggett, A. J.},
  title   = {Influence of Dissipation on Quantum Tunneling in Macroscopic Systems},
  journal = {Phys. Rev. Lett.},
  volume  = {46},
  pages   = {211},
  year    = {1981}
}

@article{HuPazZhang1992,
  author  = {Hu, B. L. and Paz, J. P. and Zhang, Y.},
  title   = {Quantum Brownian Motion in a General Environment: Exact Master Equation with Nonlocal Dissipation and Colored Noise},
  journal = {Phys. Rev. D},
  volume  = {45},
  pages   = {2843},
  year    = {1992}
}

@article{BBQ,
  title   = {Black-box superconducting circuit quantization},
  author  = {Nigg, S. E. and Paik, H. and Vlastakis, B. and Kirchmair, G. and Shankar, S. and Frunzio, L. and Devoret, M. H. and Schoelkopf, R. J. and Girvin, S. M.},
  journal = {Physical Review Letters},
  volume  = {108},
  number  = {24},
  pages   = {240502},
  year    = {2012},
  doi     = {10.1103/PhysRevLett.108.240502},
  eprint  = {1204.0587},
  archivePrefix = {arXiv},
  primaryClass  = {cond-mat.mes-hall}
}

@article{halliwell1996alternative,
  author  = {Halliwell, J. J. and Yu, T.},
  title   = {Alternative derivation of the Hu-Paz-Zhang master equation of quantum Brownian motion},
  journal = {Physical Review D},
  volume  = {53},
  pages   = {2012--2019},
  year    = {1996},
  doi     = {10.1103/PhysRevD.53.2012}
}

@article{Kubo,
  author  = {Kubo, R.},
  title   = {The Fluctuation–Dissipation Theorem},
  journal = {Rep. Prog. Phys.},
  volume  = {29},
  pages   = {255},
  year    = {1966}
}

@article{CallenWelton1951,
  author  = {Callen, H. B. and Welton, T. A.},
  title   = {Irreversibility and Generalized Noise},
  journal = {Phys. Rev.},
  volume  = {83},
  pages   = {34--40},
  year    = {1951}
}

@article{MartinSchwinger1959,
  author  = {Martin, P. C. and Schwinger, J.},
  title   = {Theory of Many-Particle Systems. I},
  journal = {Phys. Rev.},
  volume  = {115},
  pages   = {1342--1373},
  year    = {1959}
}

@article{PelargonioZaccone2023,
  author  = {Pelargonio, S. and Zaccone, A.},
  title   = {Generalized Langevin equation with shear flow and its fluctuation-dissipation theorems derived from a Caldeira-Leggett Hamiltonian},
  journal = {Phys. Rev. E},
  volume  = {107},
  pages   = {064102},
  year    = {2023},
  doi     = {10.1103/PhysRevE.107.064102},
  eprint  = {2302.03982},
  archivePrefix = {arXiv}
}

@article{CuiZaccone2018,
  author  = {Cui, B. and Zaccone, A.},
  title   = {Generalized Langevin equation and fluctuation-dissipation theorem for particle-bath systems in external oscillating fields},
  journal = {Phys. Rev. E},
  volume  = {97},
  pages   = {060102},
  year    = {2018},
  doi     = {10.1103/PhysRevE.97.060102},
  eprint  = {1802.09848},
  archivePrefix = {arXiv}
}

@article{GambaCuiZaccone2025,
  author  = {Gamba, D. and Cui, B. and Zaccone, A.},
  title   = {Open quantum systems with particle and bath driven by time-dependent fields},
  journal = {Phys. Rev. A},
  volume  = {112},
  pages   = {012207},
  year    = {2025},
  doi     = {10.1103/PhysRevA.112.012207}
}

@article{Blais2004,
  author  = {Blais, A. and Huang, R.-S. and Wallraff, A. and Girvin, S. M. and Schoelkopf, R. J.},
  title   = {Cavity quantum electrodynamics for superconducting electrical circuits: An architecture for quantum computation},
  journal = {Phys. Rev. A},
  volume  = {69},
  pages   = {062320},
  year    = {2004}
}

@article{Clerk2010,
  author  = {Clerk, A. A. and Devoret, M. H. and Girvin, S. M. and Marquardt, F. and Schoelkopf, R. J.},
  title   = {Introduction to quantum noise, measurement, and amplification},
  journal = {Rev. Mod. Phys.},
  volume  = {82},
  pages   = {1155--1208},
  year    = {2010}
}

@article{DevoretSchoelkopf2013,
  author  = {Devoret, M. H. and Schoelkopf, R. J.},
  title   = {Superconducting circuits for quantum information: An outlook},
  journal = {Science},
  volume  = {339},
  pages   = {1169--1174},
  year    = {2013}
}

@article{Liou,
  author = {von Neumann, John},
  title = {Wahrscheinlichkeitstheoretischer Aufbau der Quantenmechanik},
  journal = {Nachrichten von der Gesellschaft der Wissenschaften zu Göttingen, Mathematisch-Physikalische Klasse},
  year = {1927},
  pages = {245--272}
}

@article{FeynmanVernon1963,
  author  = {Feynman, R. P. and Vernon, F. L.},
  title   = {The Theory of a General Quantum System Interacting with a Linear Dissipative System},
  journal = {Annals of Physics},
  volume  = {24},
  pages   = {118--173},
  year    = {1963},
  doi     = {10.1016/0003-4916(63)90068-X}
}

@book{MikeIke,
  author    = {Nielsen, Michael A. and Chuang, Isaac L.},
  title     = {Quantum Computation and Quantum Information},
  edition   = {10th Anniversary Edition},
  publisher = {Cambridge University Press},
  address   = {Cambridge},
  year      = {2010},
  isbn      = {9781107002173}
}

@article{Schrodinger1926,
  author  = {Schr{\"o}dinger, E.},
  title   = {An Undulatory Theory of the Mechanics of Atoms and Molecules},
  journal = {Physical Review},
  volume  = {28},
  number  = {6},
  pages   = {1049--1070},
  year    = {1926},
  doi     = {10.1103/PhysRev.28.1049}
}

@online{Poisson,
  author       = {Cline, Douglas},
  title        = {Poisson Bracket Representation of Hamiltonian Mechanics},
  year         = {n.d.},
  organization = {LibreTexts Physics},
  url          = {https://phys.libretexts.org/Bookshelves/Classical_Mechanics/Variational_Principles_in_Classical_Mechanics_(Cline)/15%3A_Advanced_Hamiltonian_Mechanics/15.02%3A_Poisson_bracket_Representation_of_Hamiltonian_Mechanics},
  note         = {Accessed: 2026-03-08}
}

@article{Leghtas2015,
  author  = {Leghtas, Z. and Touzard, S. and Pop, I. M. and Kou, A. and Vlastakis, B. and Petrenko, A. and Sliwa, K. M. and Narla, A. and Shankar, S. and Hatridge, M. J. and Reagor, M. and Frunzio, L. and Schoelkopf, R. J. and Mirrahimi, M. and Devoret, M. H.},
  title   = {Confining the state of light to a quantum manifold by engineered two-photon loss},
  journal = {Science},
  volume  = {347},
  pages   = {853--857},
  year    = {2015}
}

@article{CaldeiraLeggett1983,
  author  = {Caldeira, A. O. and Leggett, A. J.},
  title   = {Path integral approach to quantum Brownian motion},
  journal = {Physica A},
  volume  = {121},
  pages   = {587--616},
  year    = {1983}
}

@article{Blais2021,
  author = {Blais, A. and Grimsmo, A. L. and Girvin, S. M. and Wallraff, A.},
  title = {Circuit Quantum Electrodynamics},
  journal = {Reviews of Modern Physics},
  volume = {93},
  pages = {025005},
  year = {2021},
  doi = {10.1103/RevModPhys.93.025005},
  url = {https://doi.org/10.1103/RevModPhys.93.025005}
}

@article{Arute2019,
  author = {Arute, F. et al.},
  title = {Quantum Supremacy Using a Programmable Superconducting Processor},
  journal = {Nature},
  volume = {574},
  pages = {505--510},
  year = {2019},
  doi = {10.1038/s41586-019-1666-5},
  url = {https://doi.org/10.1038/s41586-019-1666-5}
}

@article{Krantz2019,
  author = {Krantz, P. et al.},
  title = {A Quantum Engineer's Guide to Superconducting Qubits},
  journal = {Applied Physics Reviews},
  volume = {6},
  pages = {021318},
  year = {2019},
  doi = {10.1063/1.5089550},
  url = {https://doi.org/10.1063/1.5089550}
}

@article{Gambetta2017,
  author = {Gambetta, J. M. and Chow, J. M. and Steffen, M.},
  title = {Building Logical Qubits in a Superconducting Quantum Computing System},
  journal = {npj Quantum Information},
  volume = {3},
  pages = {2},
  year = {2017},
  doi = {10.1038/s41534-016-0004-0},
  url = {https://doi.org/10.1038/s41534-016-0004-0}
}

@article{Walter2017,
  author = {Walter, T. et al.},
  title = {Rapid High-Fidelity Single-Shot Dispersive Readout of Superconducting Qubits},
  journal = {Physical Review Applied},
  volume = {7},
  pages = {054020},
  year = {2017},
  doi = {10.1103/PhysRevApplied.7.054020},
  url = {https://doi.org/10.1103/PhysRevApplied.7.054020}
}

@article{Mundhada2019,
  author = {S. O. Mundhada and A. Grimm and J. Venkatraman and Z. K. Minev and S. Touzard and N. E. Frattini and V. V. Sivak and K. Sliwa and P. Reinhold et al.},
  title = {Experimental Implementation of a Raman-Assisted Eight-Wave Mixing Process},
  journal = {Physical Review Applied},
  volume = {12},
  pages = {054051},
  year = {2019},
  doi = {10.1103/PhysRevApplied.12.054051},
  url = {https://doi.org/10.1103/PhysRevApplied.12.054051}
}

@article{HuPazZhang,
  author  = {Hu, B. L. and Paz, J. P. and Zhang, Y.},
  title   = {Quantum Brownian Motion in a General Environment: Exact Master Equation with Nonlocal Dissipation and Colored Noise},
  journal = {Physical Review D},
  volume  = {45},
  pages   = {2843},
  year    = {1992}
}

@article{heisenberg1925,
  author  = {Heisenberg, W.},
  title   = {\"Uber quantentheoretische Umdeutung kinematischer und mechanischer Beziehungen},
  journal = {Zeitschrift f\"ur Physik},
  volume  = {33},
  pages   = {879--893},
  year    = {1925},
  doi     = {10.1007/BF01328377}
}

@article{born1925,
  author  = {Born, M. and Jordan, P.},
  title   = {Zur Quantenmechanik},
  journal = {Zeitschrift f\"ur Physik},
  volume  = {34},
  pages   = {858--888},
  year    = {1925},
  doi     = {10.1007/BF01328531}
}

@book{Economou2006,
  author = {E. N. Economou},
  title = {Green's Functions in Quantum Physics},
  edition = {3},
  publisher = {Springer},
  address = {Berlin},
  year = {2006},
  doi = {10.1007/3-540-28841-4}
}

@book{FetterWalecka,
  author = {A. L. Fetter and J. D. Walecka},
  title = {Quantum Theory of Many-Particle Systems},
  publisher = {Dover Publications},
  address = {New York},
  year = {2003}
}

@misc{LongQuote,
  note = {`\textbf{B. Input-Output Theory of Networks.}  While the master equation describes the system’s
damped dynamics, it provides no information on the fields
radiated by the system. Since radiated signals are what
are measured experimentally, it is of practical importance
to include those in our model.'},
  author= Walraff
}


\end{document}